\documentclass[a4]{article}
\usepackage{amssymb}
\usepackage{amsmath}
\usepackage{latexsym}
\usepackage{array}
\usepackage{bm}
\usepackage{theorem}
\usepackage{graphicx}
\usepackage{exscale}
\theoremstyle{break}
\newtheorem{Theorem}{Theorem}[section]

\def\Proof{\hfil\break{\bf Proof}\;\;\;\;}
\newtheorem{Proposition}[Theorem]{Proposition}
\newtheorem{Lemma}[Theorem]{Lemma}
\newtheorem{Corollary}[Theorem]{Corollary}

\def\hbreak{\vspace*{2mm}\hfill\break\noindent}

\def\n{\boldsymbol{n}}

\def\x{\boldsymbol{x}}
\def\y{\boldsymbol{y}}
\def\f{\boldsymbol{f}}
\def\c{\boldsymbol{c}}
\def\e{\boldsymbol{e}}
\def\mR{\mathcal{R}}

\def\Ord{\mbox{\rm Ord}}

\def\Z{\mathbb{Z}}

\def\C{\mathbb{C}}

\def\A{\boldsymbol{A}}

\def\qed{\hfill\hbox{\rule[-2pt]{3pt}{6pt}}}

\begin{document}

\title{Algebraic entropy of a multi-term recurrence of the Hietarinta-Viallet type}
\author{Ryo Kamiya$^1$, Masataka Kanki$^2$, Takafumi Mase$^1$, Tetsuji Tokihiro$^1$\\
\small $^1$ Graduate School of Mathematical Sciences,\\
\small the University of Tokyo, 3-8-1 Komaba, Tokyo 153-8914, Japan\\
\small $^2$ Department of Mathematics, Faculty of Engineering Science,\\
\small Kansai University, 3-3-35 Yamate, Osaka 564-8680, Japan}
\date{}

\maketitle


\begin{abstract}
We introduce a family of extensions of the Hietarinta-Viallet equation to a multi-term recurrence relation via a reduction from the coprimeness-preserving extension to the discrete KdV equation.
The recurrence satisfies the irreducibility and the coprimeness property although it
is nonintegrable in terms of an exponential degree growth.
We derive the algebraic entropy of the recurrence by an elementary method of calculating the degree growth.
The result includes the entropy of the original Hietarinta-Viallet equation.
\end{abstract}

\section{Introduction}
There has been a question of what is exactly the discrete integrability.
Various attempts have been made to construct a reasonable definition of discrete integrability  by analogy with that of continuous systems.
An underlying idea is that the integrability of a discrete equation is closely related to the slow growth of certain quantities.
One of the first criteria for discrete integrability is the singularity confinement test (SC test) \cite{sc}, which was introduced as a discrete analogue of the Painlev\'{e} test for differential equations.
The SC test asserts that if all the singularities of a discrete equation are resolved after a finite iterations: i.e., the information on the initial values are recovered, then the equation passes the test. 
The SC test has been successfully applied to discrete QRT mappings to discover
numerous nonautonomous recurrences including the discrete Painlev\'{e} equations \cite{sc2}.

Another famous criterion uses the algebraic entropy \cite{entropy}.
The algebraic entropy of a discrete mapping is a non-negative scalar which is related to the degree growth of the iterated mapping.
The algebraic entropy $E$ of a mapping $\varphi$ is defined as
\[
E:=\lim_{n\to\infty} \frac{\log(d_n)}{n},
\]
where $d_n$ is the degree of the $n$-th iterate $\deg \varphi^n$ of some initial condition.
If a discrete equation has zero algebraic entropy the equation is decided to be integrable, otherwise when the entropy is positive the equation is a non-integrable mapping.
In this article we hire the zero algebraic entropy condition as a working definition of discrete integrability.
It has been discovered that a certain type of discrete equations has positive algebraic entropy while passing the SC test.
The first example of this kind is the mapping by Hietarinta and Viallet \cite{hv}:
\begin{equation} \label{eq:hv}
x_{n+1}=-x_{n-1}+x_n+\frac{a}{x_n^2},
\end{equation}
where $a$ is a nonzero constant.
The algebraic entropy of \eqref{eq:hv} is derived to be $\log \frac{3+\sqrt{5}}{2}$ by an algebraic method \cite{entropy}, and by an algebro-geometric method (blowing ups and construction of a space of initial conditions) \cite{Takenawa}.

Now a lot of examples of confining equations (whose singularities are all confined) with positive algebraic entropies are known \cite{bedfordkim, redemption, redeeming, extendedhv}, however they are all equations on a one-dimensional lattice.
It has been a problem to find the singularity confining equations with exponential degree growth defined over a multi-dimensional lattice.
Recently several such examples have been discovered using the notion of the coprimeness property.
The coprimeness property was introduced as an algebraic re-interpretation of the singularity confinement test, originating from the idea that a common factor between two iterates corresponds to a common zero/pole.
Let $\varphi$ be a dynamical system of a variable $x_h$ $(h\in L)$ where $L$ is an integer lattice.
Then $\varphi$ has the coprimeness property if there exists a positive constant $D$ such that arbitrary two iterates $x_h$ and $x_k$ are pairwise coprime over the field of rational functions of the initial variables,
on condition that dist$(h,k)\ge D$, where we have introduced a non-trivial metric `dist' on the lattice $L$.
Roughly speaking the system satisfies the coprimeness property if any pair of iterates that stay far enough away from each other on the lattice is coprime.
Many of the known integrable systems satisfy the coprimeness property \cite{coprimekdv, coprime, coprimetoda}.
Moreover, many {\em non-integrable} coprimeness-preserving extensions to the well-known integrable
equations were discovered including the so-called  CP discrete KdV equation, the CP discrete Toda equation and the CP Somos-$4$ sequence \cite{extendedhv, qintegrable, JMP}.
We shall call such equations as belonging to the ``CP class'' in this article.

Let us focus on the following two-dimensional CP class equation extended from the discrete KdV equation \cite{qintegrable}:
\begin{equation}\label{eq:nonlinear}
x_{t,n}+x_{t-1,n-1}=\frac{a}{x_{t,n-1}^k}+\frac{b}{x_{t-1,n}^k}.
\end{equation}
Here $k$ is a positive even integer and $a,b\neq 0$.
Note that if $k\ge 3$ is odd the equation does not pass the singularity confinement test.
The transformation of variables corresponding to its singularity pattern is
\[
	x_{t,n} = \frac{f_{t,n} f_{t-1,n-1}}{f^k_{t-1,n} f^k_{t,n-1}}.
\]
Equation \eqref{eq:nonlinear} transforms into the following recurrence analogous to the tau-function form in the integrable cases:
\begin{equation}\label{eq:laurent}
	f_{t,n} = \frac{ - f_{t-2,n-2} f^k_{t-1,n} f^k_{t,n-1}
	+ a f^{k^2-1}_{t-1,n-1} f^{k^2}_{t,n-2} f^k_{t-1,n} f^k_{t-2,n-1}
	+ b f^{k^2-1}_{t-1,n-1} f^{k^2}_{t-2,n} f^k_{t,n-1} f^k_{t-1,n-2}
}{f^k_{t-2,n-1}f^k_{t-1,n-2}}.
\end{equation}
The irreducibility and the coprimeness of \eqref{eq:laurent} are first addressed in \cite{qintegrable} and the complete proof is published in \cite{RIMS2018}.
See Appendix for details.
In this article, we study the following recurrence
\begin{equation}\label{LHV_reduction}
x_m+x_{m-p-q}=\frac{a}{x_{m-q}^k}+\frac{b}{x_{m-p}^k} \qquad (k \in 2\Z_{>0}),
\end{equation}
where $1 \le p<q$ are positive integers coprime with each other.
Note that if $(p,q)=r$ $r\ge 2$, then the iteration splits into $r$ independent orbits on which the results in this article can be applied.
The equation \eqref{LHV_reduction} is obtained as a reduction of \eqref{eq:nonlinear} and can be considered as an extension of the Hietarinta-Viallet equation into a multi-term recurrence relation.
The equation \eqref{LHV_reduction} is transformed into the ``tau-function'' form \eqref{LHV_poly} 
\begin{align}
&f_mf_{m-2p-q}^kf_{m-p-2q}^k=-f_{m-p}^kf_{m-q}^kf_{m-2p-2q} \notag \\
&\qquad +af_{m-p}^kf_{m-p-q}^{k^2-1}f_{m-2q}^{k^2}f_{m-2p-q}^k
+bf_{m-q}^kf_{m-p-q}^{k^2-1}f_{m-2p}^{k^2}f_{m-p-2q}^k,
\label{LHV_poly}
\end{align}
via the transformation \eqref{xandf}:
\begin{equation}\label{xandf}
x_m=\frac{f_mf_{m-p-q}}{f_{m-p}^kf_{m-q}^k}.
\end{equation}
It is proved that \eqref{LHV_poly} also satisfies the  Laurent, the irreducibility and the coprimeness properties as in Appendix.

\section{Algebraic entropy}
In this section we obtain the algebraic entropy of the equation \eqref{LHV_reduction}.
It is worth noting that since \eqref{LHV_reduction} is a multi-term recurrence,
it is not easy to apply an algebro-geometric technique to obtain the space of initial conditions to derive the algebraic entropy.
Thus we stick to elementary estimation of the growth of the degrees.
From here on let us fix the set of initial variables of Equation \eqref{LHV_poly} as $\f=\{f_{-2p-2q}, f_{-2p-2q+1},...,f_{-1}\}$. The initial variables of \eqref{LHV_reduction} corresponding to $\f$ is denoted as $\x:=\{x_{-p-q},x_{-p-q+1},...,x_{-1} \}$.
Let us denote by $\Ord(r)$ the degree of a rational function $r(\x)$ with respect to $\x$: i.e.,
if we write $r=f/g$, $f,\, g \in \Z[\x,a,b]$, where
$f$, $g$ are pairwise coprime as polynomials of $\x$, then $\Ord(\x):=\max[\deg(f),\deg(g)]$.
Let us denote by $\Ord_{x_s}(f)$ the degree of $f$ with respect to
$x_s$.

The main theorem of this article gives the algebraic entropy of \eqref{LHV_reduction}.
\begin{Theorem}\label{Entropy_Theorem}
The algebraic entropy $E_{p,q}$  of the equation \eqref{LHV_reduction} for a positive even $k$ is given by
\[
E_{p,q}=\log \Lambda_{p,q},
\]
where $\Lambda_{p,q}$ is the largest real root of
\begin{equation}
\lambda^{p+q}-k(\lambda^p+\lambda^q)+1=0. \label{pq_roots}
\end{equation}
\end{Theorem}
%
%
%
Now we shall prepare several propositions.
Let us define two subsets of $\f$ as $\f_a:=\{f_{-2p-2q}, f_{-2p-2q+1},...,f_{-p-q-1}\}$ and $\f_b:=\{f_{-p-q}, f_{-p-q+1},...,f_{-1}\}$.
When we rewrite the iterate $f_m$ ($m=0,1,2,\ldots$) using $\f_a\cup \x$ or $\f_b\cup \x$ instead of $\f$, we obtain the following proposition:
\begin{Proposition} \label{prop_ugdecomp}
Each iterate $f_m$ is expressed as
\begin{subequations}
\begin{align}
f_m(\f_a,\x)&=u_m(\f_a)g_m(\x), \label{f_ug}\\
f_m(\f_b,\x)&=v_m(\f_b)h_m(\x). \label{f_vh}
\end{align}
\end{subequations}
Here $u_m,\,v_m$ are Laurent monomials that satisfy $u_mu_{m-p-q}=u_{m-p}^ku_{m-q}^k$ and $v_mv_{m-p-q}=v_{m-p}^kv_{m-q}^k$, and $g_m(\x)$, $h_m(\x)$
are irreducible Laurent polynomials that satisfy \eqref{LHV_poly}.
\end{Proposition}
\Proof
First we study the case of $-2p-2q\le m\le -1$.
By transforming the variables in $\f_b$ into those in $\f_a\cup \x$ we have
\begin{align*}
f_{-p-q}&=\frac{f_{-2p-q}^kf_{-p-2q}^k}{f_{-2p-2q}}x_{-p-q},\quad f_{-p-q+1}=\frac{f_{-2p-q+1}^kf_{-p-2q+1}^k}{f_{-2p-2q+1}}x_{-p-q+1},\cdots \\
f_{-q-1}&=\frac{f_{-p-q-1}^kf_{-2q-1}^k}{f_{-p-2q-1}}x_{-q+1},\quad
f_{-q}=\frac{f_{-2q}^kf_{-2p-q}^{k^2} f_{-p-2q}^{k^2-1}}{f_{-2p-2q}^k} x_{-q}x_{-p-q}^k, \cdots \\
f_{-1}&=\frac{f_{-p-1}^k f_{-q-1}^k}{f_{-p-q-1}}x_{-1}.
\end{align*}
Thus $f_{m}$ $(-2p-2q\le m\le -1)$ is inductively expressed as the following form:
$f_m=u_m(\f_a)g_m(\x)$ where $u_m$ is a Laurent monomial and $g_m$ is a monomial.
Note that $u_m=f_m,\,g_m=1$ ($-2p-2q \le m \le -p-q-1$).
From the transformation \eqref{xandf}, we have $u_mu_{m-p-q}=u_{m-p}^ku_{m-q}^k$.
Thus Proposition \ref{prop_ugdecomp} for $u_m,g_m$ with $m \le -1$ is proved.
Let us inductively prove the statement for $u_m$ and $g_m$ with $m\ge -1$.
Let us fix $r\ge -1$ and assume Proposition \ref{prop_ugdecomp} for all $m \le r$:
From the relations
\begin{align*}
\frac{ u_{r+1-p}^ku_{r+1-q}^ku_{r+1-2p-2q} }{ u_{r+1-2p-q}^ku_{r+1-p-2q}^k }
&=\frac{u_{r+1-p}^ku_{r+1-q}^k}{ u_{r+1-p-q} },\\
\frac{u_{r+1-p}^ku_{r+1-p-q}^{k^2-1}u_{r+1-2q}^{k^2}u_{r+1-2p-q}^k}{ u_{r+1-2p-q}^ku_{r+1-p-2q}^k }
&=\frac{u_{r+1-p}^ku_{r+1-q}^k}{ u_{r+1-p-q} },\\
\frac{u_{r+1-q}^ku_{r+1-p-q}^{k^2-1}u_{r+1-2p}^{k^2}u_{r+1-p-2q}^k}{ u_{r+1-2p-q}^ku_{r+1-p-2q}^k } 
&=\frac{u_{r+1-q}^ku_{r+1-p}^k}{ u_{r+1-p-q} },
\end{align*}
we have
\begin{align*}
f_{r+1}&=\frac{1}{f_{r+1-2p-q}^kf_{r+1-p-2q}^k }\left( -f_{r+1-p}^kf_{r+1-q}^kf_{r+1-2p-2q} \right. \\
&\quad \left. +af_{r+1-p}^kf_{r+1-p-q}^{k^2-1}f_{r+1-2q}^{k^2}f_{r+1-2p-q}^k
+bf_{r+1-q}^kf_{r+1-p-q}^{k^2-1}f_{r+1-2p}^{k^2}f_{r+1-p-2q}^k  \right)\\
&=\frac{u_{r+1-q}^ku_{r+1-p}^k}{ u_{r+1-p-q} }\frac{1}{g_{r+1-2p-q}^kg_{r+1-p-2q}^k }\left( -g_{r+1-p}^kg_{r+1-q}^kg_{r+1-2p-2q} \right. \\
&\quad \left. +ag_{r+1-p}^kg_{r+1-p-q}^{k^2-1}g_{r+1-2q}^{k^2}g_{r+1-2p-q}^k
+bg_{r+1-q}^kg_{r+1-p-q}^{k^2-1}g_{r+1-2p}^{k^2}g_{r+1-p-2q}^k  \right).
\end{align*}
Therefore we obtain
\[
u_{r+1}=\frac{u_{r+1-q}^ku_{r+1-p}^k}{ u_{r+1-p-q} },
\]
and that $g_{r+1}$ satisfies \eqref{LHV_poly}.
The irreducibility of $g_m$ follows from Theorem \ref{prop_irreducibility}.

The same argument applies to the case of $v_m(\f_b)$ and $h_m(\x)$.
\qed
\begin{Lemma}\label{lem_hx0}
$h_m(\x)$ is a polynomial of $x_{-p-q}$, whose constant term is nonzero.
\end{Lemma}
Lemma \ref{lem_hx0} is readily obtained by verifying $x_m \not\equiv 0,\, \infty $
when we substitute $x_{-p-q}=0$ into the initial variables of \eqref{LHV_reduction}.
\qed

We shall give the lower bound for the algebraic entropy in Proposition \ref{lower_bd}.
\begin{Proposition}\label{lower_bd}
Let $\Lambda_{p,q}$ be the same as in Theorem \ref{Entropy_Theorem}.
The algebraic entropy of Equation \eqref{LHV_reduction} satisfies
\[
E_{p,q} \ge \log \Lambda_{p,q}.
\]
\end{Proposition}
\Proof
Recall that $x_m=\frac{h_mh_{m-p-q}}{h_{m-p}^kh_{m-q}^k}$, where $\{h_m\}$ are pairwise coprime irreducible Laurent polynomials.
It is easy to show that $h_m$ has the following unique factorization:
\[
h_m=h_m^{(0)}h_m^{(1)},
\]
where $h_m^{(0)}$ is a Laurent monomial of $\x$, and $h_m^{(1)}$ is a polynomial of $\x$ that satisfies  $h_m^{(1)}\big|_{x_j=0} \ne 0$ for every $j$.
Let $d_m$ be the degree of $h_m$ with respect to $x_{-p-q}$.
From Lemma \ref{lem_hx0}, the term $h_m^{(0)}$ does not include $x_{-p-q}$.
Thus
\[
\Ord(x_m) \ge \Ord(h_m^{(1)}h_{m-p-q}^{(1)}) = d_m+d_{m-p-q}.
\]
Let us define a degree $d_m^*:=d_m\big|_{a=b=0}$ then $d_m \ge d_m^*$.
Since we have $x_m=x_{m-p-q}$ for $a=b=0$,
\[
h_m=\frac{-h_{m-p}^k h_{m-q}^k h_{m-2p-2q}}{h_{m-2p-q}^k h_{m-p-2q}^k},
\]
and thus $d_m^*$ satisfies
\[
d_m^*=k(d_{m-p}^*+d_{m-q}^*)-k(d_{m-2p-q}^*+d_{m-p-2q}^*)+d_{m-2p-2q}^*.
\]
Here, initial data are $d_m=d_m^*=0$ ($-2p-2q+1 \le m \le -1$), $d_{-2p-2q}=d_{-2p-2q}^*=1$.
Therefore $d_m^*$ grows as $d_m^* \sim \Lambda_{p,q}^m$, where $\Lambda_{p,q}$
 is the largest real root of
\begin{align*}
&\lambda^{2p+2q}-k(\lambda^{2p+q}+\lambda^{q+2p})+k(\lambda^p+\lambda^q)-1\\
&=(\lambda^{p+q}-1)\left(\lambda^{p+q}-k(\lambda^p+\lambda^q)+1\right)=0.
\end{align*}
Properties of $\Lambda_{p,q}$ will be discussed in Lemma \ref{Note_root} in Appendix.
Therefore using Lemma \ref{Note_sol} in Appendix, we can find a positive constant $c$
such that $d_m^* \ge c\Lambda_{p,q}^m$.
Thus we have
\[
\Ord(x_m) \ge c\Lambda_{p,q}^m,
\]
which readily derives $E_{p,q} \ge \log \Lambda_{p,q}$.
\qed
Next we obtain the upper bound for $E_{p,q}$, which is not as simple as the lower bound as is often the case with algebraic entropy of discrete equations.
\begin{Lemma} \label{gglemma}
The term $g_m$ is uniquely factorized as $g_m=g_m^{(0)}g_m^{(1)}$, where
$g_m^{(0)}$ is a monic monomial, and $g_m^{(1)}$ is a polynomial satisfying $g_m^{(1)}(\x)\big|_{x_s=0} \ne 0$ for $s=-1,-2,...,-p-q$.
Let $\alpha_s(m)$ be the degree of $g_m^{(0)}(\x)$ with respect to $x_s$,
and $\beta_s(m)$ be the degree of $g_m^{(1)}(\x)$.
Then we have the following estimation:
\begin{align}
\Ord(x_m) &\le 2 \sum_{s=-p-q}^{-1}  \left| \alpha_s(m)+\alpha_s(m-p-q)-k\alpha_s(m-p)-k\alpha_s(m-q)\right| \notag \\
&\quad +\sum_{s=-p-q}^{-1} \left| c_s(m)-kc_s(m-p)-kc_s(m-q)+c_{s}(m-p-q) \right| \notag \\
&+\sum_{s=-p-q}^{-1} k\left(c_{s}(m-p)+c_{s}(m-q)-\alpha_s(m-p)-\alpha_s(m-q)\right).
\label{main_inequality}
\end{align}
\end{Lemma}

\Proof
Recall that
\[
c_s(m):=\Ord_{x_s}(g_m)=\alpha_s(m)+\beta_s(m).
\]
Since $g_m(\x),\,g_{m-p}(\x),\,g_{m-q}(\x),\,g_{m-p-q}(\x)$ are pairwise coprime irreducible Laurent polynomials, we have
\begin{align*}
\Ord(x_m)&=\Ord\left( \frac{g_m^{(0)}g_{m-p-q}^{(0)}}{(g_{m-p}^{(0)}g_{m-q}^{(0)})^k}  \frac{g_m^{(1)}g_{m-p-q}^{(1)}}{(g_{m-p}^{(1)}g_{m-q}^{(1)})^k} \right)\\
&\le \Ord\left(\frac{g_m^{(0)}g_{m-p-q}^{(0)}}{(g_{m-p}^{(0)}g_{m-q}^{(0)})^k} \right) + \Ord\left( \frac{g_m^{(1)}g_{m-p-q}^{(1)}}{(g_{m-p}^{(1)}g_{m-q}^{(1)})^k} \right).
\end{align*}
Next, since
\[
\Ord_{x_s}\left( \frac{g_m^{(0)}g_{m-p-q}^{(0)}}{(g_{m-p}^{(0)}g_{m-q}^{(0)})^k} \right)=
\alpha_s(m)+\alpha_s(m-p-q)-k\alpha_s(m-p)-k\alpha_s(m-q),
\]
we have
\[
\Ord\left( \frac{g_m^{(0)}g_{m-p-q}^{(0)}}{(g_{m-p}^{(0)}g_{m-q}^{(0)})^k} \right) 
\le
\sum_s \left| \alpha_s(m)+\alpha_s(m-p-q)-k\alpha_s(m-p)-k\alpha_s(m-q)\right|,
\]
where the summation moves from $s=-p-q$ to $s=-1$.
Therefore,
\begin{align*}
&\Ord\left( \frac{g_m^{(1)}g_{m-p-q}^{(1)}}{(g_{m-p}^{(1)}g_{m-q}^{(1)})^k}  \right)\\
&=\max\left[\sum_s c_s(m)+c_{s}(m-p-q)-\alpha_s(m)-\alpha_s(m-p-q), \right. \\
&\qquad \qquad \left. \sum_s k\left(c_{s}(m-p)+c_{s}(m-q)-\alpha_s(m-p)-\alpha_s(m-q)\right) \right]
\end{align*}
\begin{align*}
&=\sum_s k\left(c_{s}(m-p)+c_{s}(m-q)-\alpha_s(m-p)-\alpha_s(m-q)\right) \\
&\quad + \max\left[\sum_s c_s(m)-kc_s(m-p)-kc_s(m-q)+c_{s}(m-p-q)\right. \\
&\left. \qquad\qquad -\left(\alpha_s(m)-k\alpha_s(m-p)-k\alpha_s(m-q)+\alpha_s(m-p-q)\right), \; 0 \right].
\end{align*}
Therefore we obtain \eqref{main_inequality}.
\qed
\hbreak
Following Proposition \ref{main_prop1} plays the key role in our estimation of the upper bound:
\begin{Proposition}\label{main_prop1}
For arbitrary $s$ ($-p-q \le s \le -1$), there exist positive constants $C_s,\,A_s$ such that
\[
|c_s(m)| \le C_s \Lambda_{p,q}^m,\qquad |\alpha_s(m)| \le A_s \Lambda_{p,q}^m.
\]
\end{Proposition}
Before its proof, let us complete the proof of Theorem \ref{Entropy_Theorem}.
From Proposition \ref{main_prop1} and \eqref{main_inequality} we have the upper bound for the algebraic entropy:
\begin{Corollary}\label{cor_upperbound}
We have $E_{p,q} \le \log \Lambda_{p,q}$.
\end{Corollary}
From Proposition \ref{lower_bd} and Corollary \ref{cor_upperbound} we obtain our main Theorem \ref{Entropy_Theorem}.
\qed
\hbreak

\subsection{Proof of Proposition \ref{main_prop1}}
The rest of this section is devoted to proving Proposition \ref{main_prop1}.
Let us prepare an elementary lemma on a recurrence relation.
\begin{Lemma}\label{sequence}
For a sequence $(a_m)_{m=-p-q}^\infty$, let us define $A_m:=a_m-k(a_{m-p}+a_{m-q})+a_{m-p-q}$ ($m=0,1,...$).
Suppose that there exists an integer $m_0 \in \Z_{\ge p+q}$ such that $A_m=A_{m-p-q}$ for every $m \ge m_0$.
Then there exists a positive constant $C$ such that $|a_m| \le C \Lambda_{p,q}^m$
for every $m \ge m_0$.
\end{Lemma}
\Proof
$A_m=A_{m-p-q}$ is equivalent to
\[
a_m-k(a_{m-p}+a_{m-q})+k(a_{m-2p-q}+a_{m-p-2q})-a_{m-2p-2q}=0.
\]
The characteristic polynomial of this linear recurrence is
\begin{align*}
&\lambda^{2p+2q}-k(\lambda^{p+2q}+\lambda^{2p+q})
+k(\lambda^{q}+\lambda^{p})-1\\
&=(\lambda^{p+q}-1)\left\{(\lambda^{p+q}-k(\lambda^q+\lambda^p)+1\right\},
\end{align*}
whose largest root with respect to the absolute value is $\Lambda_{p,q}$ from Note \ref{Note_root}.
Therefore there exists a constant $C>0$ such that $|a_m| \le C \Lambda_{p,q}^m$.
\qed
\hbreak
%
%
%
\subsubsection{The case of $p=1$, $q=2$:}
First let us prove the former inequality on $c_s(m)$.
We have
\begin{align*}
g_{-6}&=g_{-5}=g_{-4}=1,\\
g_{-3}&=x_{-3},\ g_{-2}=x_{-3}^kx_{-2},\ g_{-1}=x_{-3}^{k^2+k}x_{-2}^kx_{-1}.
\end{align*}
Let us denote by $y_m$ the following degree: $y_m:=c_{-3}(m)=\Ord_{x_{-3}} (g_m)$.
By calculating $\y=(y_{-6},y_{-5},y_{-4},...)$ we have
\[
y_{-6}=y_{-5}=y_{-4}=0,\quad y_{-3}=1,\;y_{-2}=k,\;y_{-1}=k^2+k.
\]
For $m \ge 0$ we have
\[
g_m=\frac{-g_{m-1}^kg_{m-2}^kg_{m-6}+ag_{m-1}^kg_{m-3}^{k^2-1}g_{m-4}^{k^2+k}
+bg_{m-2}^{k^2+k}g_{m-3}^{k^2-1}g_{m-5}^k}{g_{m-4}^kg_{m-5}^k}.
\]
By comparing the degrees of the three terms on the right hand side
\begin{align*}
y_m^{(1)}&:=k(y_{m-1}+y_{m-2})-k(y_{m-4}+y_{m-5})+y_{m-6},\\
y_m^{(2)}&:=ky_{m-1}+(k^2-1)y_{m-3}+k^2y_{m-4}-ky_{m-5},\\
y_m^{(3)}&:=(k^2+k)y_{m-2}+(k^2-1)y_{m-3}-ky_{m-4},
\end{align*}
we obtain
\begin{equation}
y_m=\max[y_m^{(1)},y_m^{(2)},y_m^{(3)}], \label{ym123}
\end{equation}
unless an unexpected cancellation occurs.
Precisely speaking, if the two terms, for example $g_{m-1}^kg_{m-2}^kg_{m-6}$
and $ag_{m-1}^kg_{m-3}^{k^2-1}g_{m-4}^{k^2+k}$ has the same degree with respect to
all of $x_s$ ($s=-1,-2,-3$)，and the degree is greater than that of the third term，it is possible that the degree satisfies $y_m<\max[y_m^{(1)},y_m^{(2)},y_m^{(3)}]$.
This type of cancellation is inductively proved to be impossible later in this proof.

Let us define a sequence $Y_m:=y_m-k(y_{m-1}+y_{m-2})+y_{m-3}$. Then
\[
Y_{-3}=1,\quad Y_{-2}=0,\quad Y_{-1}=0.
\]
We shall prove that $Y_m=Y_{m-3}$ for every $m$.
Let us define
\[
Y_m^{(i)}:=y_m^{(i)}-k(y_{m-1}+y_{m-2})+y_{m-3}\qquad (i=1,2,3).
\]
Then we have
\[
Y_m^{(1)}=Y_{m-3},\qquad Y_m^{(2)}=-kY_{m-2},\qquad Y_m^{(3)}=-kY_{m-1},
\]
and
\begin{equation}\label{Y_12_123}
Y_m=\max[Y_{m-3},-kY_{m-2},-kY_{m-1}].
\end{equation}
We shall use the notation (applicable only in this section)  $Y_m=a_{I}$ to denote that $Y_m=a$ and the maximum/maxima in Equation \eqref{Y_12_123} is attained on the $i$th term(s) for all $i\in I$. For example, $Y_0=\max[1,0,0]=1$ where maximum is attained on the first term $1$, and thus we write $Y_0=1_{1}$. 
The successive iterations give
\[
Y_0=1_{1},\quad Y_1=\max[0,0,-k]=0_{1,2},\quad Y_2=0_{1,3},\quad Y_3=1_{1},\quad Y_4=0_{1,2},\quad Y_5=0_{1,3}, \cdots,
\]
on condition that no unexpected cancellation occurs in \eqref{Y_12_123}.

Next we study $c_{-2}(m)$ and $c_{-1}(m)$.
Note that $\Ord_{x_{-2}}g_m=y_{m-1},\, \Ord_{x_{-1}}g_m=y_{m-2}$.
We redefine $y_{m}:=c_{-2}(m), y_{m}^{(i)} (i=1,2,3)$ and $Y_{m}=y_{m}-k(y_{m-1}+y_{m-2})+y_{m-3}$ (we use the same symbols $y_m$ and $Y_m$ as $s=-3$ to ease notation). Then $Y_m$ satisfies Equation~\eqref{Y_12_123}, with the initial condition
\[
Y_{-3}=0,\; Y_{-2}=1,\,Y_{-1}=0 \quad (s=-2).
\]
If a cancellation does not occur, we have
\[
Y_0=0_{1,3},\quad Y_1=1_{1},\quad Y_2=0_{1,2},\quad Y_3=0_{1,3},\quad Y_4=1_{1}, \quad Y_5=0_{1,2}, \cdots.
\]
For $s=-1$ let us redefine $y_{m}:=c_{-1}(m), y_{m}^{(i)} (i=1,2,3)$ and $Y_{m}=y_{m}-k(y_{m-1}+y_{m-2})+y_{m-3}$. Then we have
\[
Y_{-3}=0,\; Y_{-2}=0,\,Y_{-1}=1,
\]
\[
Y_0=0_{1,2},\quad Y_1=0_{1,3},\quad Y_2=1_{1},\quad Y_3=0_{1,2},\quad Y_4=0_{1,3}, \quad Y_5=1_{1}, \cdots.
\]
From these results, for any $m$, there exists at least one $s\in\{-1,-2,-3\}$ such that the right hand side of \eqref{Y_12_123} attains its maximum only for one term (i.e., only one subscript in our notation).
Thus the degrees of $y_m^{(i)}$ $(i=1,2,3)$ are all distinct from each other as rational functions of $\{x_{-1},x_{-2},x_{-3}\}$.
Therefore it is proved inductively that no unexpected cancellation occurs while iterating \eqref{Y_12_123} (and thus \eqref{ym123}) and that
we have $Y_m-Y_{m-3}=0$ for all $m\ge 0$.
Thus, from Lemma \ref{sequence}, there exists a constant $C_s>0$ such that
$|y_m| \le C_s \Lambda_{1,2}^m$ for  all $s=-1,-2,-3$.
Now $|c_s(m)|\le C_s \Lambda_{1,2}^m$ ($s=-1,-2,-3$) is proved.

Next let us prove the latter inequality on $\alpha_s(m)$.
From Lemma \ref{gglemma}, $g_m$ has the factorization $g_m=g_m^{(0)}g_m^{(1)}$.
Let $z_m$ be the degree of $g_m^{(0)}$ with respect to $x_{-3}$: i.e., $z_m:=\alpha_{-3}(m)$.
It is clear that $z_m=y_m$ ($-6 \le m \le -1$).
For $m\ge 0$, $z_m$ is iteratively defined as
\begin{equation} \label{zm123}
z_m=\min[z_m^{(1)},z_m^{(2)},z_m^{(3)}],
\end{equation}
where we use three auxiliary variables as
\begin{align*}
z_m^{(1)}&:=k(z_{m-1}+z_{m-2})-k(z_{m-4}+z_{m-5})+z_{m-6},\\
z_m^{(2)}&:=kz_{m-1}+(k^2-1)z_{m-3}+k^2z_{m-4}-kz_{m-5},\\
z_m^{(3)}&:=(k^2+k)z_{m-2}+(k^2-1)z_{m-3}-kz_{m-4}.
\end{align*}
Equation \eqref{zm123} is true unless we encounter a non-trivial cancellation of terms just like in $y_m$.
In order to avoid unexpected cancellations, it is sufficient that at least one of ``degrees of monomial parts of $g_m$ are distinct (discussion on $Z_m$ below)'' or ``degrees of $g_m$ are distinct (discussion on $Y_m$)'' is satisfied.
Let $Z_m:=z_m-k(z_{m-1}+z_{m-2})+z_{m-3}$.
We have $Z_{-3}=1,\,Z_{-2}=Z_{-1}=0$.
Let us define
\[
Z_m^{(i)}:=z_m^{(i)}-k(z_{m-1}+z_{m-2})+z_{m-3}\qquad (i=1,2,3).
\]
Then
\[
Z_m^{(1)}=Z_{m-3},\qquad Z_m^{(2)}=-kZ_{m-2},\qquad Z_m^{(3)}=-kZ_{m-1},
\]
and thus
\begin{equation}\label{z_12_123}
Z_m=\min_i[Z_m^{(i)}]=\min[Z_{m-3},-kZ_{m-2},-kZ_{m-1}].
\end{equation}
It is easy to check that
\[
Z_0=0_{2,3},\quad Z_i=0_{1,2,3} \ (1 \le i).
\]

Let us study the case of $s=-2$.
Let us abuse the notation and redefine $z_m=\alpha_{-2}(m)$ and so on.
Then Equation \eqref{z_12_123} is satisfied with the initial condition $Z_{-3}=0,\,Z_{-2}=1,\,Z_{-1}=0$. Thus $Z_m$ is periodic with period three for $m \ge 3$ as
\[
Z_0=-k_{2},\quad Z_1=0_{2},\quad Z_2=0_{1,3},\quad Z_3=-k_{1}, \quad Z_4=0_{1,2},\quad Z_5=0_{1,3}, \cdots.
\]
The same discussion applies to $\alpha_{-1}(m)$.
The redefined $Z_m$ satisfies Equation \eqref{z_12_123} with the initial condition
$Z_{-3}=0,\,Z_{-2}=0,\,Z_{-1}=1$. Thus $Z_m$ is again periodic with period three for $m \ge 3$ as
\[
Z_0=-k_{3},\quad Z_1=-k_{2},\quad Z_2=1_{1},\quad Z_3=-k_{1,3}, \quad Z_4=-k_{1,2},\quad Z_5=1_{1}, \cdots.
\]
In the case of $m\equiv 0,2\mod 3, m\le 4$, there exists at least one $s\in\{-1,-2\}$ such that the right hand side of \eqref{z_12_123} attains its minimum only for one term.
In the case of $m\equiv 1\mod 3, m\le 4$, the degrees of the first two terms of the right hand side of $g_m$ are zero with respect to $x_{-2}$ and $x_{-3}$,
they must have the following form:
\[
x_{-1}^{K} G_1+ax_{-1}^{K}\ G_2.
\]
Here $K=z_m^{(1)}=z_m^{(2)}$ and $G_1,G_2$ are irreducible.
On the other hand, from the discussion of $Y_m$,
their degrees satisfy $\Ord(G_1)\neq \Ord (G_2)$, and therefore their highest order terms cannot be cancelled out.
Therefore from Lemma \ref{sequence}, there exists a constant $A_{s}>0$ such that
$|z_m| \le A_s \Lambda_{1,2}^m$ for any $s=-1,-2,-3$.
Thus $|\alpha_s(m)|\le A_s \Lambda_{1,2}^m$ ($s=-1,-2,-3$) is proved.
\qed
%
%
\subsubsection{The case of $p=1$, $q \ge 3$:}

By successive iterations we have
\begin{align*}
g_s&=1\quad (-2q-2 \le s \le -q-2),\quad
g_{-q-1}=x_{-q-1},\quad
g_{-q}=x_{-q-1}^kx_{-q},\cdots, \\
g_{-2}&=x_{-q-1}^{k^{q-1}}x_{-q}^{k^{q-2}}\cdots x_{-3}^kx_{-2},\quad
g_{-1}=x_{-q-1}^{k^q+k}x_{-q}^{k^{q-1}}\cdots x_{-2}^kx_{-1}.
\end{align*}
Let $y_m^{(s)}:=c_{s}(m)=\Ord_{x_{s}} (g_m)$ for $s=-q-1,\cdots, -1$.
For example,
\[
y_s^{(-q-1)}=0\quad (-2q-2 \le s \le -q-2),\, y_{-q-1}^{(-q-1)}=1,\,y_{-q}^{(-q-1)}=k,\cdots , y_{-2}^{(-q-1)}=k^{q-1},\,y_{-1}^{(-q-1)}=k^q+k.
\]
Just like the case of $p=1,\,q=2$, let us define
\begin{align*}
y_{m,1}^{(s)}&:=k(y_{m-1}^{(s)}+y_{m-q}^{(s)})-k(y_{m-2-q}^{(s)}+y_{m-1-2q}^{(s)})+y_{m-2-2q}^{(s)},\\
y_{m,2}^{(s)}&:=ky_{m-1}^{(s)}+(k^2-1)y_{m-1-q}^{(s)}+k^2y_{m-2q}^{(s)}+ky_{m-2-q}^{(s)}-k(y_{m-2-q}^{(s)}+y_{m-1-2q}^{(s)}),\\
y_{m,3}^{(s)}&:=ky_{m-q}^{(s)}+(k^2-1)y_{m-q-1}^{(s)}+k^2y_{m-2}^{(s)}+ky_{m-1-2q}^{(s)}-k(y_{m-2-q}^{(s)}+y_{m-1-2q}^{(s)}).
\end{align*}
Let
$Y_m^{(s)}:=y_m^{(s)}-k(y_{m-1}^{(s)}+y_{m-q}^{(s)})+y_{m-q-1}^{(s)}$ and $Y_{m,i}^{(s)}:=y_{m,i}^{(s)}-k(y_{m-1}^{(s)}+y_{m-q}^{(s)})+y_{m-q-1}^{(s)}$ for $i=1,2,3$.
Then
\[
Y_{m,1}^{(s)}=Y_{m-1-q}^{(s)},\qquad Y_{m,2}^{(s)}=-kY_{m-q}^{(s)},\qquad Y_{m,3}^{(s)}=-kY_{m-1}^{(s)}.
\]
Therefore
\begin{equation}\label{initial_Y1q}
Y_m^{(s)}=\max\left[ Y_{m-1-q}^{(s)},\, -kY_{m-q}^{(s)},\, -kY_{m-1}^{(s)}\right].
\end{equation}
Let us study the case of $s=-q-1$ first.
We have
\[
(Y_{-q-1}^{(-q-1)},Y_{-q}^{(-q-1)},Y_{-q+1}^{(-q-1)},\cdots, Y_{-1}^{(-q-1)})=(1,0,0,...,0).
\]
Therefore if there is no unexpected cancellation of terms, which shall be proved inductively in the course of the proof, we have
\begin{align*}
Y_0^{(-q-1)}&=\max[1,-k\cdot 0,-k \cdot 0]=1_1,\quad
Y_1^{(-q-1)}=\max[0,-k \cdot 0, -k \cdot 1]=0_{1,2}, \\
Y_j^{(-q-1)}&=0_{1,2,3}\ (2\le j\le q-1),\quad
Y_{q}^{(-q-1)}=\max[Y_{-1}^{(-q-1)}, -k Y_0^{(-q-1)}, -k Y_{q-1}^{(-q-1)}]=0_{1,3},\\
Y_{q+1}^{(-q-1)}&=\max[Y_0^{(-q-1)}, -kY_1^{(-q-1)},-kY_q^{(-q-1)}]=1_1.
\end{align*}
Therefore $Y_m^{(-q-1)}$ has period $q+1$.

Next we study the case of $s=-q$.
Initial values of \eqref{initial_Y1q} are
\[
(Y_{-q-1}^{(-q)},Y_{-q}^{(-q)},Y_{-q+1}^{(-q)},\cdots ,Y_{-1}^{(-q)})=(0,1,0,...,0).
\]
The iterations give
\begin{align*}
Y_0^{(-q)}&=0_{1,3},\quad
Y_1^{(-q)}=1_1,\quad
Y_2^{(-q)}=0_{1,2},\quad
Y_j^{(-q)}=0_{1,2,3}\ (3\le j\le q), \\
Y_{q+1}^{(-q)}&=\max[Y_0^{(-q)}, -kY_1^{(-q)},-kY_q^{(-q)}]=0_{1,3}.
\end{align*}
Therefore $Y_m^{(-q)}$ is also periodic with the period $q+1$.
The same discussion shows that $Y_m^{(s)}$ has period $q+1$ for all $-q+1 \le s \le -1$.
When we fix $s$, the right hand side of Equation \eqref{initial_Y1q} attains its maximum only for one term if $m=s$ (mod $q+1$).
Therefore no irregular cancellation is proved to be impossible.
From Lemma \ref{sequence}, for each $s$ there exists a constant $C_s>0$ such that $|c_s(m)| \le C_s\Lambda_{1,q}^m$.

Next let us study $\alpha_s(m)$ $(-q-1 \le s \le -1)$.
Let $z_m^{(s)}=\alpha_s(m)$ and
$Z_m^{(s)}:=z_m^{(s)}-k(z_{m-1}^{(s)}+z_{m-q}^{(s)})+z_{m-q-1}^{(s)}$.
If no cancellation occurs we have
\begin{equation} \label{zmequationq3}
Z_m^{(s)}=\min[Z_{m-q-1}^{(s)},-kZ_{m-q}^{(s)},-kZ_{m-1}^{(s)}],
\end{equation}
where the initial values are $Z_s^{(s)}=1$, $Z_j^{(s)}=0\, (j \ne s)$.
Let us prove that no cancellation occurs using a procedure similar to the $(p,q)=(1,2)$ case.
In the case of $s=-1$ we have:
\begin{align*}
Z_0^{(-1)}&=\min[0,0,-k]=-k_3,\quad Z_1^{(-1)}=\min[0,0,k^2]=0_{1,2},\quad
Z_{j}^{(-1)}=0_{1,2,3}\ (2\le j\le q-2),\\
 Z_{q-1}^{(-1)}&=\min[0,-k,0]=-k_2,\quad
Z_q^{(-1)}=\min[1,k^2,k^2]=1_1,\quad Z_{q+1}^{(-1)}=\min[-k,0,-k]=-k_{1,3}, \\
Z_{q+2}^{(-1)}&=\min[0,0,k^2]=0_{1,2}, \quad Z_{j}^{(-1)}=0_{1,2,3}\ (q+3 \le j \le 2q-2),\quad
Z_{2q-1}^{(-1)}=\min[0,k^2,0]=0_{1,3},\\
Z_{2q}^{(-1)}&=\min[-k,-k,0]=-k_{1,2}, \quad
Z_{2q+1}^{(-1)}=\min[1,k^2,k^2]=1_{1}, \cdots.
\end{align*}
Therefore $Z_m^{(-1)}$ has period $q+1$.

For $-2 \le s \le -q$, we have $Z_m^{(s)}=-k$ for $m\equiv q+s$ (mod $q+1$), which is the unique minimum in the right hand side of Equation \eqref{zmequationq3}.
Otherwise $Z_m^{(s)}=0$. Therefore $Z_m^{(s)}=Z_{m-q-1}^{(s)}$.
For $s=-q-1$ we have $Z_{-q-1}^{(-q-1)}=1, Z_m^{(-q-1)}=0\, (-q\le m \le -1)$ and
\[
Z_0^{(-q-1)}=\min[1,0,0]=0_{2,3}.
\]
We have $Z_m^{(-q-1)}=0$ for any $m\ge 0$. Thus $Z_m^{(-q-1)}=Z_{m-q-1}^{(-q-1)}$.

From the discussion above, for $m\equiv 0,1,2,...,q-2$ (mod $q+1$),
$Z_{m}^{(m-q)}$ has the unique minimum in the right hand side of Equation \eqref{zmequationq3}.
For $m\equiv q$, $Z_m^{(-1)}$  has the unique minimum.
For $m \equiv q-1$,
monomials of $\x\setminus\{ x_{-1} \}$ do not appear in the first and the second terms of $g_m$.
Therefore the cancellation is impossible, taking into account the fact that these two terms have distinct degrees from the discussion of $Y_m$.
Therefore, for each $s$, there exists a constant $A_s>0$ such that
$|\alpha_s(m)| \le A_s \Lambda_{1,q}^m$.
\qed

%
%
\subsubsection{The case of $q>p\ge 2$:}

Since $p$ and $q$ are coprime, let us write $lp<q<(l+1)p$, $r=q-lp$ where $0<r<p$. 
We shall use the same notations as previous parts.
Let us define $y_m^{(s)}:=c_{s}(m)$ for $s=-q-p,\cdots, -1$.
First let us derive $y_m^{(-q-p)}$.
Values of $y_m^{(-q-p)}$ for $-q-p \le m \le -1$ are
\begin{align}
&y_{-q-p}=1,\ y_{-q-p+i}=0\ (i=1,2,\cdots, p-1), \notag \\ 
&y_{-q+tp}=k^{t+1},\ y_{-q+tp+i}=0\ (i=1,2,\cdots, p-1,\, t=0,1,\cdots, l-1,\, (i,t)\neq (r,l-1)),\notag \\
& y_{-p}(=y_{-q+(l-1)p+r})=k, \notag \\ 
&y_{-r}=k^{l+1},\ y_j=0\ (j=-r+1,\cdots,-1), \label{ymgeneralcase}
\end{align}
where we have omitted the superscripts ${}^{(-p-q)}$ for simplicity.
For example, when $(q,p)=(17,5), k=2$, we have
\[
(y_{-22}^{(-22)},\cdots, y_{-1}^{(-22)})=
(1,0,0,0,0,2,0,0,0,0,4,0,0,0,0,8,0,2,0,0,16,0).
\]
In the case of $s=-p-q+1$, we have
\[
y_{-q-p}^{(-q-p+1)}=0,\ y_{m+1}^{(-q-p+1)}=y_m^{(-q-p)}\, (m=-q-p,\cdots,-2),
\]
which is derived by shifting the sequence \eqref{ymgeneralcase} to the right and adding $0$ to the left.
In a similar manner we have
\[
y_m^{(s)}=0\, (-q-p\le m\le s-1),\ y_{m+s+q+p}^{(s)}=y_m^{(-q-p)}\, (-q-p\le m\le -q-p-s-1),
\]
for $-q-p+2\le s\le -1$.
In particular, $(y_{-q-p}^{(-1)},\cdots, y_{-1}^{(-1)})=(0,0,....,0,1)$.
Note that  $y_m^{(s)}=0$ ($m \le -p-q-1$) for any $s$.
Let
\[
Y_m^{(s)}=y_m^{(s)}-k(y_{m-p}^{(s)}+y_{m-q}^{(s)})+y_{m-p-q}^{(s)}.
\]
Then, for a fixed $s$, we have $Y_s^{(s)}=1$，$Y_j^{(s)}=0$ ($-p-q \le j \le -1,\, j \ne s$).
Since we have
\begin{equation} \label{Ymgeneral}
Y_m^{(s)}=\max[Y_{m-p-q}^{(s)},-kY_{m-q}^{(s)}, -kY_{m-p}^{(s)}],
\end{equation}
it is inductively proved that
$Y_m^{(-p-q+j)}=1$ ($m\equiv j\!\mod p+q$),
$Y_m^{(-p-q+j)}=0$ ($m \not \equiv j\!\mod p+q$) for $j=0,1,...,p+q-1$.
In the case of $Y_m^{(s)}=1$, only the first term of the right hand side of \eqref{Ymgeneral} gives the maximum, and therefore there is no unexpected cancellation of the terms.
Thus we have $Y_m^{(s)}=Y_{m-p-q}^{(s)}$ for all $s$.

Next let us investigate $z_m^{(s)}=\alpha_s (m)$ and $Z^{(s)}_m:=z_m^{(s)}-k(z_{m-p}^{(s)}+z_{m-q}^{(s)})+z_{m-p-q}^{(s)}$ for $-p-q \le s$.
First $Z_m^{(s)}=Y_m^{(s)}$ for $-p-q \le s \le -1$.
We readily obtain
\begin{equation} \label{Zmgeneral}
Z_m^{(s)}=\min [ Z_{m-p-q}^{(s)}, -k Z_{m-q}^{(s)}, -k Z_{m-p}^{(s)} ].
\end{equation}
It is proved in a similar manner to the previous case that $Z_m^{(s)}$ is periodic with respect to $m$ with period $p+q$. The sketch is as follows: for a fixed $-p-q+1 \le s \le -1$ we have
\[
Z_{2p+q+s}^{(s)}=-k,\ Z_{p+2q+s}^{(s)}=-k,\ Z_{2p+2q+s}^{(s)}=1,\ \mbox{otherwise}\ Z_m^{(s)}=0,
\]
for $p+q\le m\le 2p+2q-1$,
and this sequence continues periodically with period $p+q$.
Moreover $Z_{2p+q+s}^{(s)}$ has the unique minimum $1$ on the right hand side of \eqref{Zmgeneral}.
Therefore no cancellation is possible.
For $s=-p-q$, we have $Z_m^{(-p-q)}=0$ for all $m$, and this case neither denies a cancellation.
The proof of Proposition~\ref{main_prop1} is now complete.
\qed

\section{Further generalizations}
Let us give a further generalization to Equation \eqref{LHV_reduction} via a reduction from the higher-dimensional lattice equation.
Here is a higher dimensional analogue of \eqref{eq:nonlinear}:
\begin{equation}
x_{t+1,\n}+x_{t-1,\n}=\sum_{i=1}^d \left( \frac{a_i}{x_{t,\n+\e_i}^{k_i}}+\frac{b_i}{x_{t,\n-\e_i}^{l_i}}\right)
\qquad (k_i, l_i \in 2\Z_+),
\label{pedKdV_eq}
\end{equation}
where each $\e_i\in\mathbb{Z}^d$ ($i=1,2,\cdots ,d$) is the unit vector $(0,\cdots,0,1,0,\cdots,0)$ whose $i$th component is $1$, and $\n=\sum_{i=1}^d n_i\e_i$ denotes a point on the lattice.
The set of initial variables are taken from those on $t=0$ and $t=1$ hyperplanes
and evolve the equation towards $t \ge 2$.
Equation \eqref{pedKdV_eq} is proved to satisfy the coprimeness property  \cite{highHV} (the exact statement is that ``two iterates $x_{t,\n}$ and $x_{t',\n'}$ are coprime in $\mathbb{Q}\left( \{x_{0,\n},x_{1,\n}\}_{\n\in\mathbb{Z}^d}, \{a_i,b_i\}_{i=1}^d \right)$ on condition that $|t-t'|>2$ or $|\n-\n'|>2$'') under the following condition:
\begin{equation} \label{conditionhighHV}
\min_{1\le i \le d} [k_i m_i-1]>\max_{1\le i \le d} [k_i m_i].
\end{equation}
Let us give one of the reductions of \eqref{pedKdV_eq} to one-dimensional lattice systems. Let $d=2, \n=(n,m)$ and $p>q>r$ be positive integers.
Suppose that $x_N:=x_{t,n,m}$ is constant
if we fix one $N=pt+qn+rm$.
Then we have the following recurrence relation:
\begin{equation}
x_{N+p}+x_{N-p}=\frac{a_1}{x_{N+q}^{k_1}}+\frac{b_1}{x_{N-q}^{m_1}}+\frac{a_2}{x_{N+r}^{k_2}}+\frac{b_2}{x_{N-r}^{m_2}}. \label{eqqqrr}
\end{equation}
It is conjectured from several examples that, when the condition \eqref{conditionhighHV} is satisfied, the dynamical degree of \eqref{eqqqrr}
is equal to the largest real root of
\[
\lambda^{2p}-k_1 \lambda^{p+q}-k_2\lambda^{p+r}-m_2\lambda^{p-r}-m_1\lambda^{p-q}+1=0,
\]
which is the ``characteristic'' polynomial of its singularity structure.
As for the second order systems (three-term recurrences), the relation between the degree growth and the singularity structures are well investigated. See
\cite{Halburd, Exp, Mase2018} for details.
By taking $p=5/2, q=3/2, r=1/2$, $k_1=m_2=4, k_2=m_1=2$ and shifting $N = n-5/2$, we have
\begin{equation}
x_n+x_{n-5}=\frac{1}{x_{n-4}^2}+\frac{1}{x_{n-3}^4}+\frac{1}{x_{n-2}^2}+\frac{1}{x_{n-1}^4}. \label{eq2424}
\end{equation}
From a numerical experiment the dynamical degree of Equation \eqref{eq2424} is estimated to be in $(4.63551,4.63552)$, while the largest real root of 
\[
\lambda^5-4\lambda^4-2\lambda^3-4\lambda^2-2\lambda+1=0
\]
is $4.6355149\cdots$.
Here we have used the Diophantine calculation \cite{Diophantine} for our estimation:
the height of an iterate as a rational number is calculated instead of the its degree.
The height $H(r)$ of a non-zero rational number $r=\frac{p}{q}$, where $p,q$ are pairwise coprime integers, is defined as $H(r)=\max(|p|,|q|)$ and serves as the arithmetic complexity of rationals.
When we take arbitrary rational numbers as the initial variables, then every iterate $x_n\in\mathbb{Q}$. The speed of the growth of $\log H(x_n)$ is conjectured to be equal to that of $\deg x_n$.
Precisely speaking, the following limit
\[
\lim_{n\to\infty}\frac{\log H(x_{n+1})}{\log H(x_n)}
\]
is conjectured to converge to the dynamical degree of the mapping.
Another example is
\[
x_n+x_{n-6}=\frac{1}{x_{n-5}^2}+\frac{1}{x_{n-4}^2}+\frac{1}{x_{n-2}^2}+\frac{1}{x_{n-1}^2},
\]
whose dynamical degree is estimated to be in $(2.82320, 2.82322)$.
This quantity is close to $2.8232019\cdots$, which is the largest real root of
\[
\lambda^6-2\lambda^5-2\lambda^4-2\lambda^2-2\lambda+1=0.
\]
On the other hand, if we study
\[
x_n+x_{n-6}=\frac{1}{x_{n-5}^2}+\frac{1}{x_{n-3}^4}+\frac{1}{x_{n-1}^2},
\]
which does not satisfy \eqref{conditionhighHV}, the estimation of its dynamical degree is in $(2.61832,2.61835)$, while the root of
\[
\lambda^6-2\lambda^5-4\lambda^3-2\lambda+1=0
\]
is $\lambda=2.618{\bf 0}339\cdots$.
The discrepancy between these values seems to be beyond a numerical error and serves as a counter-example for the conjecture without \eqref{conditionhighHV}.
These are only {\em conjectural} topics, however we wish to give rigorous results in future correspondences.

\section{Conclusion}
In this article we have introduced a recurrence relation \eqref{LHV_reduction} through a reduction from the coprimeness-preserving extension to the discrete KdV equation \eqref{eq:nonlinear}.
Equation \eqref{LHV_reduction} also satisfies the irreducibility and the coprimeness property and is considered as one generalisation of the Hietarinta-Viallet equation to a multi-term recurrence.
As the main Theorem \ref{Entropy_Theorem} we have derived that the algebraic entropy of \eqref{LHV_reduction} is given by the largest real root of the polynomial related to the singularity pattern of the equation.
Although the proof is slightly complicated when obtaining the upper bound of the entropy, only elementary tools have been used.
Finally we have introduced a higher-dimensional lattice equation \eqref{pedKdV_eq}.
We have given several numerical simulations of the algebraic entropies of reduced mappings of \eqref{pedKdV_eq} and have conjectured a property similar to Theorem \ref{Entropy_Theorem}.

\section*{Acknowledgement}
This work is supported by JSPS KAKENHI grant numbers 17K14211, 18H01127 and 18K13438.

\appendix

\section{Supplementary materials}
\subsection{Review on the coprimeness of Equation \eqref{eq:laurent} and \eqref{LHV_poly}}
Let us review the results on the coprimeness property of the tau-function form of the coprimeness-preserving discrete KdV equation \eqref{eq:laurent} and its reduction \eqref{LHV_poly}.
\begin{Theorem}[\cite{qintegrable, RIMS2018}] \label{thm:laurent} 
Let $R$ be a unique factorization domain (UFD) and let $a, b \in R$ be nonzero.
Then, Equation \eqref{eq:laurent} has the Laurent property on any good domain, i.e.\ every iterate is a Laurent polynomial of the initial variables on any good domain.
Moreover, every iterate is irreducible as a Laurent polynomial.
\end{Theorem}
Here a nonempty subset $H \subset \mathbb{Z}^2$ is a good domain (with respect to Equation \eqref{eq:laurent}) if it satisfies the following two conditions \cite{rims}:
\begin{itemize}
\item
If $(t, n) \in H$, then $(t+1, n), (t, n+1) \in H$.
\item
For any $h \in H$,  $\#\{ h' \in H \mid h' \le h \}<\infty$,
where
we denote by ``$\le$'' the product order on the lattice $\mathbb{Z}^2$: i.e.,
$h \le h' \Leftrightarrow
t \le t'$ and $n \le n'$
for $h = (t, n), h' = (t', n') \in \mathbb{Z}^2$.
\end{itemize}
Note that the first quadrant is a good domain with the initial variables on the $L$-shaped area $\{(t,n)\, |\, t=0,1,\, n\ge 0\ \mbox{or}\ n=0,1,\, t\ge 0\}$.
It is proved that the reduction \eqref{LHV_poly} also satisfies the Laurent, the irreducibility and the coprimeness properties.
\begin{Theorem}[\cite{RIMS2018}]\label{prop_irreducibility}
Let us denote by $\f$ the set of initial variables of \eqref{LHV_poly}. Then, for every iterate $f_m$ we have
\[
f_m \in \mR:=\Z\left[\f^{\pm},\, a,\, b \right].
\]
Moreover each iterate is irreducible and arbitrary two iterates are pairwise coprime in $\mR$.
\end{Theorem}
Proof of Theorem \ref{prop_irreducibility} is explained in \cite{RIMS2018} (Japanese article).
From the discussion in \cite{investigation}, if a multi-dimensional lattice equation has the Laurent property on any good domain, then its reductions to lower-dimensional lattices preserve the Laurent property.
Therefore, the Laurentness of \eqref{LHV_poly} follows from the Theorem \ref{thm:laurent} on the two-dimensional lattice equation \eqref{eq:laurent}.
However, the irreducibility and the coprimeness do not trivially follow by the reduction and we need to prove them inductively with respect to $m$.
In the case of $(p,q)=(1,2)$ the induction process is not very different from that of the extended Hietarinta-Viallet equation \cite{extendedhv}, however when $p\ge 2$ or $q\ge 3$ the calculation is a bit complicated.
First we show Lemma on the factorization of the Laurent polynomials, whose proof is given in \cite{coprime}:
\begin{Lemma}[\cite{coprime}, Lemma~2]\label{lem:toki}
Let $R$ be a UFD and $\{p_1,p_2,\cdots,p_m\}$ and $\{q_1,q_2,\cdots ,q_m\}$ be two sets of independent variables satisfying  for $j=1,2,\cdots, m$ the following properties:
\[
p_j\in R\left[ q_1^{\pm}, q_2^{\pm},\cdots ,q_m^{\pm}\right], \qquad q_j\in R\left[ p_1^{\pm}, p_2^{\pm},\cdots ,p_m^{\pm}\right],
\]
and that $q_j$ is irreducible as an element of $R\left[ p_1^{\pm}, p_2^{\pm},\cdots ,p_m^{\pm}\right]$.
Let $f$ be an irreducible Laurent polynomial
\[
f(p_1,\cdots,p_m)\in R\left[ p_1^{\pm}, p_2^{\pm},\cdots ,p_m^{\pm}\right],
\]
and let $g$ be another Laurent polynomial
\[
g(q_1,\cdots, q_m) \in R\left[ q_1^{\pm}, q_2^{\pm},\cdots ,q_m^{\pm}\right],
\]
where $f(p_1,\cdots,p_m)=g(q_1\cdots, q_m)$.
Then the function $g$ is factorized as
\[
g(q_1,\cdots, q_m)=p_1^{r_1}p_2^{r_2}\cdots p_m^{r_m}\, \tilde{g}(q_1,\cdots,q_m),
\]
where $r_1,r_2, \cdots, r_m\in\mathbb{Z}$ and $\tilde{g}(q_1,\cdots,q_m)$ is irreducible in $R \left[ q_1^{\pm}, q_2^{\pm},\cdots ,q_m^{\pm}\right]$.
\end{Lemma}
\subsubsection{The case of $(p,q)=(1,2)$}
When $(p,q)=(1,2)$ the equation \eqref{LHV_poly} is
\begin{equation}\label{eq:redp1q2}
	f_m = \frac{-f^k_{m-1} f^k_{m-2} f_{m-6} + a f^k_{m-1} f^{k^2-1}_{m-3} f^{k^2+k}_{m-4}
	+ b f^{k^2+k}_{m-2} f^{k^2-1}_{m-3} f^k_{m-5}}{f^k_{m-4} f_{m-5}^k},
\end{equation}
whose initial variables are $\f=\{f_j\}_{j=-6}^{-1}$.
Here are two Lemmata for the proof:
\begin{Lemma}\label{lem:tpolyp1q2}
Each $f_m$ ($m \ge 0$) is a polynomial of $f_{-6}$, whose constant term is nonzero.
\end{Lemma}
\begin{Lemma}\label{lem:specialp1q2}
Let us substitute the following values in Equation \eqref{eq:redp1q2}:
\[
	a = b = 0, \quad
	f_{-6} = t, \quad
	f_{-5} = \cdots = f_{-1} = 1.
\]
Then $f_m$ has a form $f_m = \pm t^{\alpha_m}$, where $\alpha_m$ is given by
\begin{align*}
	\alpha_m &= k(\alpha_{m-1} + \alpha_{m-2} - \alpha_{m-4} + \alpha_{m-5}) + \alpha_{m-6} \quad (m \ge 0), \\
	\alpha_{-6} &= 1, \quad
	\alpha_{-5} = \cdots = \alpha_{-1} = 0.
\end{align*}
In particular we have $\alpha_m > (k - 1)\alpha_{m-1}$ for $m \ge 6$.
\end{Lemma}
Now let us show Theorem \ref{prop_irreducibility} for $(p,q)=(1,2)$, using an induction with respect to $m$.
\paragraph{Step~$1$: irreducibility of $f_m$ $(m=0,1,2)$}

The iterate $f_0$ is linear with respect to $f_{-6}$, and thus is irreducible (from here on the irreducibility is considered in $\mR=R[\{f_j^{\pm}\}_{j=-6}^{-1}]$ unless otherwise stated) and not invertible.
Using Lemma \ref{lem:toki} for $R=\mathbb{Z}[a,b]$, $\{p_j\}_{j=1}^6=\{f_j\}_{j=-5}^0$, $\{q_j\}_{j=1}^6=\{f_j\}_{j=-6}^{-1}$, $f_1$ is expressed as
\[
	f_1 = F_1\prod_{j=-5}^0 f_j^{r_j},
\]
where $F_1$ is irreducible and each $r_j$ is an integer.
Since $f_1$ is a Laurent polynomial and $f_0$ is irreducible and non-invertible, we must have $r_0\ge 0$.
On the other hand we have
\[
	f_1 \equiv \frac{b f^{k^2+k}_{-1} f^{k}_{-2}}{f^k_{-3}} \mod f_0.
\]
Since $f_0$ is linear with respect to $f_{-6}$, the iterate $f_0$ cannot divide $f_1$, and therefore $r_0 = 0$.
Moreover, $r_j$ $(-5\le j\le -1)$ are units in $\mR$, thus $f_1$ is irreducible and is coprime with $f_0$.
From the irreducibility of $f_1$ in $\mR$, the iterate $f_2$ is trivially irreducible in $R[\{f_j^{\pm}\}_{j=-5}^0]$ by shifting all the subscripts.
Thus using Lemma \ref{lem:toki}, we have $f_2=f_0^{r_2} F_2$, where $F_2$ is irreducible in $\mR$ and $r_2$ is a non-negative integer. 
We can show from a simple computation that $f_2 \not\equiv 0\!\mod f_0$.
Thus $r_2=0$ and $f_2$ is irreducible.

\paragraph{Step~$2$: irreducibility of $f_m$ $(m=3,4,5,6)$}

By a similar discussion to the previous step, we can inductively show the irreducibility of $f_j$ and express $f_j=f_0^{r_j} F_j$ for $j=3,4,5,6$, where $F_j$ is irreducible and $r_j$ is a non-negative integer.
The case of $j=3$ is easy since it is readily obtained that $f_3\not\equiv 0\!\mod f_0$.
Let us prove the irreducibility of $f_4$. By a direct calculation we have
\[
	f_2 = \frac{\left( -f^k_0 f_{-4} + a f^{k^2+k}_{-2} f^{k^2-1}_{-1} \right) f^k_1 + \mathcal{O}(f^{k^2+k}_0)}{f^k_{-2} f^k_{-3}}, \quad
	f_3 = \frac{- f_{-3} f^k_2 f^k_1 + \mathcal{O}(f^{k^2+k}_0)}{f^k_{-1} f^k_{-2}},
\]
and
\begin{align*}
	- f^k_3 f^k_2 f_{-2} + b f^{k^2+k}_2 f^{k^2-1}_1 f^k_{-1}
	&\equiv \frac{f^{k^2-1}_1 f^{k^2+k}_2}{f^{k^2}_{-1} f^{k^2-1}_{-2}} \left( -f_1 f^k_{-3} + b f^{k^2+k}_{-1} f^{k^2-1}_{-2} \right) \mod f^{k+1}_0 \\
	&\equiv \frac{f^{k^2-1}_1 f^{k^2+k}_2}{f^{k^2}_{-1} f^{k^2-1}_{-2}} \left( \frac{f^k_2 f_{-2} - a f^{k^2+k}_0 f^{k^2-1}_{-2}}{f^k_{-4}} f^k_0 \right)  \mod f^{k+1}_0.
\end{align*}
Thus,
\begin{align*}
	f_4 &= \frac{-f^k_3 f^k_2 f_{-2} + b f^{k^2+k}_2 f^{k^2-1}_1 f^k_{-1} + \mathcal{O}(f^{k^2+k}_0)}{f^k_0 f^k_{-1}}\\
	 \equiv& \frac{f^{k^2-1}_1 f^{k^2+k}_2}{f^{k^2+k}_{-1} f^{k^2-1}_{-2} f^{k}_{-4}} \left( f^k_{-1} f_{-5} - a f^{k^2+k}_{-3} f^{k^2-1}_{-2} \right) \not \equiv 0 \mod f_0.
\end{align*}
Therefore $f_4$ is irreducible.

\noindent
The case of $f_5$ is done as follows: by a direct calculation we have
\[
	-f^k_4 f^k_3 f_{-1} + a f^k_4 f^{k^2-1}_2 f^{k^2+k}_1 \equiv \frac{f^k_4 f^{k^2-1}_2 f^{k^2+k}_1 f^k_0 f_{-4}}{f^{k^2-1}_{-1} f^{k^2+k}_{-2}} \mod f^{k+1}_0,
\]
and thus
\begin{align*}
	f_5 &\equiv \frac{f^k_4 f^{k^2-1}_2 f^{k^2}_1 f_{-4}}{f^{k^2-1}_{-1} f^{k^2+k}_{-2}} + b \frac{f^{k^2+k}_3 f^{k^2-1}_2}{f^k_1} \mod f_0 \\
	&= \frac{f^{k^2-1}_2}{f^{k^2-1}_{-1} f^{k^2+k}_{-2} f^k_1} \left( f^k_4 f^{k^2+k}_1 f_{-4} + b f^{k^2+k}_3 f^{k^2-1}_{-1} f^{k^2+k}_{-2} \right).
\end{align*}
It is sufficient to prove that $F := f^k_4 f^{k^2+1}_1 f_{-4} + b f^{k^2+k}_3 f^{k^2-1}_{-1} f^{k^2+k}_{-2} \not \equiv 0 \mod f_0$.
By choosing the initial values $f_{-6} = a + b, f_{-5} = \cdots = f_{-1} = 1$, and by taking the parameters as $a > 1$, $b > 0$, we can show that $F>0$.
Thus $F$ is not divisible by $f_0$.
Therefore $f_5$ is irreducible.

\noindent
The irreducibility of $f_6$ is proved in a similar manner, since we can prove that
\[
	f_6 \equiv \frac{f^{k^2-1}_3}{f^k_2 f^k_1} \left(a f^k_5 f^{k^2+k}_2 + b f^{k^2+k}_4 f^k_1 \right) \not \equiv 0 \mod f_0.
\]

\paragraph{Step~$3$: coprimeness of $f_m$ $(0\le m\le 6)$}

Suppose that $f_i$ and $f_j$ ($i > j$) are not pairwise coprime, then we can express $f_i = u f_j$ where $u$ is an invertible element in $\mR$.
On the other hand, from Lemma \ref{lem:specialp1q2}, $u$ must include the factor $t^{\alpha_i - \alpha_j}$.
Therefore the constant term of $f_j$ as a polynomial of $t$ must be zero, which contradicts Lemma \ref{lem:tpolyp1q2}.

\paragraph{Step~$4$: irreducibility and coprimeness of $f_m$ $(m\ge 7)$}

From Lemma \ref{lem:toki}, $f_7$ is factorized in two ways as
\[
	f_7 = f^{r_0}_0 F = f^{r_1}_1 \cdots f^{r_6}_6 F',
\]
where $F,F'$ are irreducible in $\mR$ and $r_j$ are non-negative integers.
Suppose that $f_7$ is not irreducible, then the factorization is limited to the form
$f_7 = u f_0 f_j$ where $u$ is invertible and $j\in\{1, \cdots, 6\}$.
Thus we have
\[
	\alpha_7 > (k - 1) \alpha_6 + 1 \ge \alpha_j + \alpha_3,
\]
which contradicts Lemma \ref{lem:tpolyp1q2}.
The case of $m\ge 8$ can be done in the same manner.
The pairwise coprimeness of $f_i$ and $f_j$ $(i>j)$ is proved in exactly the same manner as in Step~$3$.
The proof of Theorem \ref{prop_irreducibility} for $(p,q)=(1,2)$ is now complete.
\qed
\subsubsection{The case of $p=1$ and $q\ge 3$}
Next let us prove the case of $p = 1$ and $q \ge 3$.
It is worth noting that $p, q, 2p, p + q, 2q, 2p + q, p + 2q$ are all distinct when $(p,q)\neq (1,2)$.
Let us present four Lemmata, which are applicable for every $(p,q)$ with $p\ge 1$ and $q\ge 2$.
\begin{Lemma}\label{lem:polyab0}
For every $s \ge 0$, $f_s$ is a polynomial of $f_{-2p-2q}, f_{-2p-2q+1}, \ldots, f_{-p-2q-1}$ whose constant term is nonzero.
The statement is true even when we substitute $a=0$ or $b=0$.
\end{Lemma}
\begin{Lemma}\label{lem:irredcs}
Let $c^{(j)}_s$ be the degree of $f_s|_{a=b=0}$ with respect to $f_{-2p-2q+j}$ ($j = 0, \ldots, p - 1$)
and let  $c_s = \left( c^{(0)}_s, \ldots, c^{(p-1)}_s \right)$.
Then we have the following properties:
\begin{enumerate}
\item[(i)]
$c^{(j)}_s = c^{(0)}_{s-j}$ for every $s \ge j$.

\item[(ii)]
If we have $f_s = u \prod_{j \in J} f_j$ where $u$ is invertible, then $c_s = \sum_{j \in J} c_j$.

\item[(iii)]
Let $f_s$ and $f_r$ be irreducible Laurent polynomials.
If $c_s \ne c_r$, $f_s$ and $f_r$ are pairwise coprime.
\end{enumerate}
\end{Lemma}

\Proof
(i) is trivial since we have
\[
	f_s|_{a=b=0} = -\frac{f^k_{s-p} f^k_{s-q} f_{s-2p-2q}}{f^k_{s-2p-q} f^k_{s-p-2q}}.
\]

\noindent
(ii)
Since $u$ is invertible, $u$ does not depend on $a, b$.
From Lemma \ref{lem:polyab0}, $u$ does not depend on $f_{-2p-2q}, \ldots, f_{-p-2q-1}$.
By substituting $a=b=0$ into $f_s = u \prod_{j \in J} f_j$, we obtain $c_s = \sum_{j \in J} c_j$.

\noindent
(iii)
Suppose that $f_s$ and $f_r$ are both irreducible but not coprime with each other.
Then there exists an invertible element $u$ such that $f_s = u f_r$.
Thus we have $c_s = c_r$ from (ii).
\qed

\begin{Lemma}\label{lem:iredipjq}
The iterate $f_s$ is irreducible if either $s \not \in \{ip + jq\, |\, i, j \in \mathbb{Z}_{\ge 0}\}$ or $0 \le s \le p + q$ is satisfied.
\end{Lemma}

\Proof
\paragraph{Step~$1$}

In the case of $s = 0, \ldots, p - 1$,
$f_s$ is linear with respect to $f_{-2p-2q+s}$ whose constant term is nonzero.
Thus $f_s$ is irreducible and is not a unit.
Note that $f_s$ does not depend on the initial variables $f_{-2p-2q+i}$ ($0 \le i \le p - 1, i \ne s$).

\paragraph{Step~$2$}

In the case of $s \ne ip + jq$ ($i, j \in \mathbb{Z}_{\ge 0}$),
$f_m$($m \ge 0$) depends on $f_{-2p-2q}$ if and only if $m$ can be written as $m = ip + jq$ where $i, j$ are nonnegative integers.
Thus $f_s$ does not depend on $f_{-2p-2q}$.
From Lemma \ref{lem:toki}, by assuming the irreducibility of $f_m$ for every $m\le s-1$,
$f_s$ can be factorized as $f_s = f^r_0 F$, where $r$ is a nonnegative integer and $F$ is irreducible.
Since $f_s$ is independent of $f_{-2p-2q}$ we must have $r = 0$.
Thus $f_s$ is irreducible.

\paragraph{Step~$3$}

In the case of $1 \le s \le p + q$,
let us define $g_s$ as the value of $f_s$ where we substitute the following values
\begin{align*}
	f_{-2p-2q} &= \frac{f^{k^2-1}_{-p-q}}{f^k_{-p} f^k_{-q}} \left(a f^k_{-p} f^{k^2}_{-2q} f^k_{-2p-q} + b f^k_{-q} f^{k^2}_{-2p} f^k_{-p-2q} \right), \\
	f_m &< 0 \quad (-2p - 2q \le m \le -p - q), \\
	f_m &> 0 \quad (-p - q + 1 \le m \le -1),
\end{align*}
into the initial variables.
Suppose that $a,b>0$.
It is clear that $g_0 = 0$ and $g_s$ satisfies
\[
	g_s = \frac{-g_{s-2p-2q} g^k_{s-p} g^k_{s-q} + \left( a g^k_{s-p} g^{k^2}_{s-2q} g^k_{s-2p-q} + b g^k_{s-q} g^{k^2}_{s-2p} g^k_{s-p-2q} \right) g^{k^2-1}_{s-p-q}}{g^k_{s-p-2q} g^k_{s-2p-q}}.
\]
Therefore $g_s > 0$.
By the same argument as in Step~$2$, we conclude that $f_s$ is irreducible.
\qed

\begin{Lemma}\label{lem:cpq}
Let us prepare two functions $C_{p,q}, \widetilde{C}_{p,q}$ by
\[
	C_{p,q} = \frac{f^{k^2-1}_{p} f^{k^2}_{q} f^k_{2p}}{f^{k^2-1}_{-q} f^{k^2}_{-p} f^k_{-2q} f^k_{p-q}}, \quad
	\widetilde{C}_{p,q} = \frac{f^{k^2}_p f^{k^2-1}_q f^k_{2q}}{f^{k^2-1}_{-p} f^{k^2}_{-q} f^k_{-2p} f^k_{-p+q}}.
\]
Then we have
\begin{align*}
	f_{2p+q} &\equiv C_{p,q} \left(f^k_{p-q} f_{-p-2q} - a f^{k^2}_{p-2q} f^k_{-p-q} f^{k^2-1}_{-q} \right) + \frac{a f^k_{p+q} f^{k^2}_{2p-q} f^{k^2-1}_p}{f^k_{p-q}} \mod f_0, \\
	f_{p+2q} &\equiv \widetilde{C}_{p,q} \left(f^k_{-p+q} f_{-2p-q} - b f^{k^2}_{-2p+q} f^k_{-p-q} f^{k^2-1}_{-p} \right) + \frac{b f^k_{p+q} f^{k^2}_{-p+2q} f^{k^2-1}_q}{f^k_{-p+q}} \mod f_0.
\end{align*}
\end{Lemma}
%
%
Now let us begin the proof of Theorem \ref{prop_irreducibility} for $p = 1, q \ge 3$.
Equation \eqref{LHV_poly} is
\[
	f_{s} = \frac{f^k_{s-1} f^k_{s-q} \left( -f_{s-2-2q} \right) + \left( a f^k_{s-1} f^{k^2}_{s-2q} f^k_{s-2-q} + b f^k_{s-q} f^{k^2}_{s-2} f^k_{s-1-2q} \right) f^{k^2-1}_{s-1-q}}{f^k_{s-1-2q} f^k_{s-2-q}}.
\]
From Lemma \ref{lem:iredipjq}, $f_s$ is irreducible for $0 \le s \le q + 1$.
Let $g_s$ be the values of the iterates $f_s$ when we substitute
\[
	f_m = \begin{cases}
		2 & (m = -2q - 2) \\
		-1 & (m = -2q) \\
		1 & (-2q - 1 \le m \le -1, m \ne -2q)
	\end{cases}
\]
into the initial variables of $f_s$ and take $a = b = 1$.
Clearly $g_0 = 0$.

\paragraph{Step~$1$}

It is readily obtained that if $g_s \ne 0$, then $f_s$ is irreducible:
From Lemma \ref{lem:toki}, there exist a nonnegative integer $r$ and an irreducible element $F$ such that $f_s = f^r_0 F$.
If we assume that $r > 0$, then $g_s = 0$.
From here on we shall prove $g_s \ne 0$ by a direct calculation.
The first few terms are
\[
	g_s = \begin{cases}
		g^{k^2}_{s-2} & (3 \le s \le q - 1) \\
		g^k_{s-1} & (s = q) \\
		-g^k_{s-1} & (s = q + 1)
	\end{cases}.
\]
Therefore
\[
	g_s = \begin{cases}
		1 & (3 \le s \le q - 1, s \ \text{is odd}) \\
		2^{k^{s-2}} & (3 \le s \le q - 1, s \ \text{is even}) \\
		g^k_{q-1} & (s = q) \\
		-g^{k^2}_{q-1} & (s = q + 1)
	\end{cases}.
\]
When $q$ is odd  we have
\[
	g_{q-1} = 1, \quad
	g_q = 1, \quad
	g_{q+1} = -1,
\]
and when $q$ is even we have
\[
	g_{q-1} = 2^{k^{q-3}}, \quad
	g_q = 2^{k^{q-2}}, \quad
	g_{q+1} = -2^{k^{q-1}}.
\]

\paragraph{Step~$2$: irreducibility of $f_s$:}

In the case of $s = q + 2$, from Lemma \ref{lem:cpq} we have
\[
	g_{q+2} = g^{k^3}_{q-1} \ne 0.
\]
In the case of $q + 3 \le s \le 2q - 1$,
we have
\[
	g_{q+r+2} = \frac{  -g^k_{s-1} g^k_{s-q} + \left( g^k_{s-1} g^k_{s-q-2} + g^k_{s-q} g^{k^2}_{s-2} \right) g^{k^2-1}_{s-1-q}}{g^k_{s-2-q}},
\]
for $1 \le r \le q - 3$.
Thus
\[
	g_{q+r+2} = \begin{cases}
		\left( N_r - 1 \right) g^k_{q+r+1} + N_r g^{k^2}_{q+r} & (r\ \text{is odd}) \\
		\left( 1 - N_r \right) g^k_{q+r+1} + N_r g^{k^2}_{q+r} & (r\ \text{is even})
	\end{cases},
\]
where
\[
	N_r = 2^{k^{r-1}(k^2-1)}.
\]
Therefore we have $g_{q+r+2} > 0$ if $r$ is odd, and $g_{q+r+2} < 0$ if $r$ is even.
These inequalities are clear when $r$ is odd. When $r$ is even, we can prove this by
\begin{align*}
	g_{q+r+2} &= -(N_r - 1) \left( (N_{r-1} - 1) g^k_{q+r} + N_{r-1} g^{k^2}_{q+r-1} \right)^k + N_r g^{k^2}_{q+r} \\
	&< -\left( (N_r - 1) (N_{r-1} - 1)^k - N_r \right) g^{k^2}_{q+r}.
\end{align*}
Therefore we have $g_s \ne 0$ for $q + 3 \le s \le 2q - 1$.

\noindent
In the case of $s=2q$, since
\[
	f_{2q} \equiv \frac{-f^k_{2q-1} f^k_q f_{-2} + b f^k_q f^{k^2}_{2q-2} f^k_{-1} f^{k^2-1}_{q-1}}{f^k_{-1} f^k_{q-2}} \mod f_0,
\]
we have
\[
	g_{2q} = \frac{g^k_q}{g^k_{q-2}} \left( -g^k_{2q-1} + g^{k^2-1}_{q-1} g^{k^2}_{2q-2} \right).
\]
It is sufficient to prove that $G = -g^k_{2q-1} + g^{k^2-1}_{q-1} g^{k^2}_{2q-2}\neq 0$.
If $q$ is even, since $q \ge 4$ and $g_{q-1} = 1$, we must have $g_{2q-1} = \pm g^k_{2q-2}$ in order to achieve $G = 0$.
This is not possible because
\[
	g_{2q-1} = \left( N_{q-3} - 1 \right) g^k_{2q-2} + N_{q-3} g^{k^2}_{2q-3} > g^k_{2q-2}.
\]
If $q$ is odd, since $g_{q-1} = 2^{k^{q-3}}$ and thus $g^{k^2-1}_{q-1} = N_{q-2}$, we have
\[
	G = -g^k_{2q-1} + N_{q-2} g^{k^2}_{2q-2}.
\]
In the case of $q = 3$, $G \ne 0$ is obtained by a direct calculation.
In the case of $q \ge 5$, we must have $g_{2q-1} = \pm N_{q-3} g^k_{2q-2}$ when $G = 0$.
However, since $g_{2q-1} = (1-N_{q-3}) g^k_{2q-2} + N_{q-3} g^{k^2}_{2q-3}$
whose right hand side is negative, we must have $g_{2q-1} =- N_{q-3} g^{k}_{2q-2}$,
and therefore $g^k_{2q-2} =- N_{q-3} g^{k^2}_{2q-3}$, which is not possible.
Thus $G \neq 0$.

\noindent
In the case of $s=2q+1$, from Lemma \ref{lem:cpq} we have
\[
	g_{2q+1} = g^{k^2-1}_q g^k_{2q} \left( g^k_{q-1} - g^{k^2}_{q-2} \right) + \frac{g^k_{q+1} g^{k^2}_{2q-1} g^{k^2-1}_q}{g^k_{q-1}}.
\]
If $q$ is odd, $g_{2q+1} > 0$ is readily obtained since $g_{q-2} = 1$, $g_{q-1} = 2^{k^{q-3}}$.
If $q$ is even, we have 
\[
	g_{q+r+2} \equiv 1 \mod 3,
\]
for $2 \le r \le q - 3$.
Since we have $g^k_{q-1} - g^{k^2}_{q-2} \equiv 0\! \mod 3$ and that none of the four iterates $g_{q+1}, g_{2q-1}, g_q, g_{q-1}$ is divisible by $3$, we have
\[
	g_{2q+1} \not \equiv 0 \mod 3.
\]
Thus $g_{2q+1} \ne 0$.

\noindent
In the case of $s = 2q + 2$, $g_{2q+2} \ne 0$ is obtained from
\[
	g_{2q+2} = \frac{g^{k^2-1}_{q+1}}{g^k_1 g^k_{q}} \left( g^k_{2q+1} g^{k^2}_{2} g^k_{q} + g^k_{q+2} g^{k^2}_{2q} g^k_{1} \right).
\]
The proof of the irreducibility of $f_s$ $(s\ge 2q+3)$ is omitted.

\paragraph{Step~$3$: coprimeness of $f_s$:}

Let us prove that $f_s$ and $f_r$ are pairwise coprime if $s > r \ge 0$.
The degrees $c^{(0)}_j$ in Lemma \ref{lem:irredcs} satisfy
\begin{align*}
	c^{(0)}_0 &= c^{(0)}_{-2q-2} = 1, \quad
	c^{(0)}_{-1} = \cdots = c^{(0)}_{-2q-1} = 0, \\
	c^{(0)}_j &= k \left( c^{(0)}_{j-1} + c^{(0)}_{j-q} - c^{(0)}_{j-2-q} - c^{(0)}_{j-1-2q} \right) + c^{(0)}_{j-2-2q} \quad (j \ge 1).
\end{align*}
Thus we have $c^{(0)}_{j} \ge k c^{(0)}_{j-1}$ for every $j \ge -2q - 1$ inductively.
Therefore $c^{(0)}_{s} > c^{(0)}_{r}$. From Lemma \ref{lem:irredcs} (iii), $f_s$ and $f_r$ are coprime with each other.
\qed


\subsubsection{The case of $p = 2$}
When $p=2$, $q$ must be odd.
Let $g_s$ be the values of the iterates $f_s$ when we substitute
\begin{equation}
	f_m = \begin{cases}
		2 & (m = -2p -2q ) \\
		-1 & (m = -2q) \\
		1 & (-2p - 2q+1 \le m \le -1, m \ne -2q)
	\end{cases} \label{initialp23}
\end{equation}
into the initial variables of $f_s$ and take $a = b = 1$.
Clearly $g_0=0$.
Our goal is to prove that $g_s \ne 0$ for every $s \ge 1$.
The discussion goes in a similar manner to the case of  $p = 1$, $q \ge 3$.

\paragraph{Step~$1$: irreducibility of $f_s$ $(1\le s\le 2q+4)$:}
A direct calculation shows that
\[
	g_s = \begin{cases}
		N_i & (s = 4i) \\
		1 & (s \ne 4i)
	\end{cases}
\]
for every $1 \le s \le q - 1$, where we have defined $N_i = 2^{k^{2(i-1)}}$.
Calculating further we have
\[
	g_q = g^k_{q-2} = 1, \quad
	g_{q+1} = g^{k^2}_{q-3}, \quad
	g_{q+2} = -g^k_q = -1, \quad
	g_{q+3} = g^{k^2}_{q-1},
\]
where we have used the fact that $q$ is odd.
From Lemma \ref{lem:cpq} we have
\[
	g_{q+4} = g^k_{q+2} g^{k^2}_{-q+4} g^{k^2-1}_2 g^{-k}_{2-q} =1 \ne 0.
\]
Thus we have $g_s \ne 0$ for $1 \le s \le q + 4$.

\noindent
Next we show $g_s \ne 0$ for $q + 5 \le s \le 2q - 1$.
By using $r=s-q-4$, we have
\[
	g_{q+r+4} =\left\{ \left( -g^k_{r+4} + g^{k^2}_{r+4-q} g^k_r g^{k^2-1}_{r+2} \right) g^k_{q+r+2} + g^k_{r+4} g^{k^2-1}_{r+2} g^{k^2}_{q+r} \right\} g^{-k}_r,
\]
and therefore,
\[
	g_{q+5} = g^{k^2}_{q+1}, \quad
	g_{q+6} = -1 + 2^{k^2-1} + 2^{k^2-1} = 2^{k^2}-1 = N_2 - 1, \quad
	g_{q+7} = g^{k^2}_{q+3}.
\]
Inductively we have the following expression for $3 \le r \le q - 5$:
\[
	g_{q+r+4} = \begin{cases}
		\left( -\dfrac{N^k_{i+1}}{N^k_i} + 1 \right) g^k_{q+r+2} + \dfrac{N^k_{i+1}}{N^k_i} g^{k^2}_{q+r} & (r = 4i) \\
		g^{k^2}_{q+r} & (r = 4i + 1, 4i + 3) \\
		\left( N^{k^2-1}_{i+1} - 1 \right) g^k_{q+r+2} + N^{k^2-1}_{i+1} g^{k^2}_{q+r} & (r = 4i + 2)
	\end{cases}.
\]
If $r$ is odd, it is clear that $g_{q+r+4} \ne 0$.
If $r$ is even, $g_{q+r+4} \equiv g_{q+r+2} \mod 2$ and thus inductively
$g_{q+r+4} \equiv 1 \mod 2 \neq 0$.
Lastly we need to prove that $g_s \ne 0$ for $2q\le s \le 2q+4$ one by one.
The case of $s=2q$ is readily obtained from $g_{2q} = -g^k_{2q-2} + g^{k^2}_{2q-4}$ and the fact that $q$ is odd.
We also have $g_{2q+1}  \equiv 1 \mod 2$ by a direct calculation.
In the case of $s=2q+2$, using the second equation in Lemma \ref{lem:cpq}, we have
$g_{q+2} = -1$, $g_q = g_{q-2} = g_{q-4} = 1$ and
$g_{2q+2} = g^k_{q+2} g^{k^2}_{2q-2} g^{k^2-1}_q g^{-k}_{q-2} = g^{k^2}_{2q-2} \ne 0$.
The cases of $s=2q+3,2q+4$ are omitted since they are proved by a direct calculation.

\paragraph{Step~$2$: coprimeness of $f_s$ $(0\le s\le 2p+4)$:}
The variables $c_j^{(0)}$ in Lemma \ref{lem:irredcs}  satisfy
\begin{align*}
	c^{(0)}_0 &= c^{(0)}_{-2q-4} = 1, \quad
	c^{(0)}_{-1} = \cdots = c^{(0)}_{2q-3} = 0, \\
	c^{(0)}_j &= k \left( c^{(0)}_{j-2} + c^{(0)}_{j-q} - c^{(0)}_{j-4-q} - c^{(0)}_{j-2-2q} \right) + c^{(0)}_{j-4-2q} \quad (j \ge 1).
\end{align*}
Thus we have $c^{(0)}_{2i} = k^{i} \, (0 \le i \le q-1)$ and
\begin{align*}
	c^{(0)}_{2i+1} &= 0 \, \left(0 \le i \le \frac{q - 3}{2} \right), \
	c^{(0)}_q = k, \
	c^{(0)}_{q+2i} =  (i + 1) k^{i+1} - (i - 1) k^{i-1} \ (1 \le i \le q-1), \\
	c^{(0)}_{2q} &= k^q + k^2, \quad
	c^{(0)}_{2q+2} = k^{q+1} + 3 k^3 - k, \quad
	c^{(0)}_{2q+4} =  k^{q+2} + 6 k^4 - 4 k^2 + 1.
\end{align*}
Therefore $c_s \ne c_r$ for $0\le s<r\le 2p+4$.
From (iii) of Lemma \ref{lem:irredcs}, $f_s$ and $f_r$ must be pairwise coprime.

\paragraph{Step~$3$: irreducibility and coprimeness of $f_s$ $(2q+5\le s)$:}

Let us first prove the irreducibility of $f_s$ for $s \ge 2q + 5$.
From the previous steps, if we suppose that $f_s$ is not irreducible then there exist
a reversible element $u$ and $1 \le r \le 2q + 4$ such that $f_s = u f_0 f_r$.
Therefore from (ii) of Lemma \ref{lem:irredcs} we must have $c_s = c_0 + c_r$, and thus
$c^{(0)}_s = c^{(0)}_r + 1, \ c^{(0)}_{s-1} = c^{(0)}_{r-1}$.
However, this leads us to a contradiction since we can prove that $c_s\neq c_0+c_r$ as follows: for $0 \le i \le q + 2$
we have $c^{(0)}_{2i} > c^{(0)}_{2i-1}$, $c^{(0)}_{i+2} \ge k c^{(0)}_{i}$ and $c^{(0)}_{i+q} \ge k c^{(0)}_{i}$.
If $s \ge 2q + 5$ and $s$ is even, we have $c^{(0)}_s \ge k c^{(0)}_{2q+4}$. Thus $c^{(0)}_s \ne c^{(0)}_r + 1$. If $s$ is odd we have $c^{(0)}_{s-1} \ge c^{(0)}_{2q+4}$ and thus $c^{(0)}_{s-1} \ne c^{(0)}_{r-1}$.

The coprimeness of $f_s$ and $f_r$ $(0\le s\le r)$ is proved in a similar manner.
\qed


\subsubsection{The case of $p \ge 3$}
When $p \ge 3$, every pair from	$\{ip, q + ip, 2q + ip\, |\,  (i = 0, 1, 2, \ldots)\}$
is distinct from each other.
Since $p$ and $q$ are pairwise coprime, we have the following Lemma \ref{lem:rpqipjq}.

\begin{Lemma}\label{lem:rpqipjq}
For $m, n \in \mathbb{Z}_{\ge 0}$ let us define $s = mp + nq$.
Then $s$ is uniquely expressed as
\[
	s = rpq + ip + jq \quad
	(r \in \mathbb{Z}_{\ge 0}, 0 \le i \le q - 1, 0 \le j \le p - 1).
\]
Moreover, this expression maps $s$ to $(r,i,j)$ bijectively. 
\end{Lemma}
From Lemma \ref{lem:iredipjq}, $f_s$ is irreducible if $0 \le s \le p + q$ or $s \ne ip + jq$ ($i, j \in \mathbb{Z}_{\ge 0}$).
Let $g_s$ be the same value as in \eqref{initialp23} in the case of $p=2$.
Our goal is to prove the irreducibility of $f_s$ for every $s\ge 1$.
Basically we have only to prove that $g_s \ne 0$, however, for $s=2q$ we have $g_{2q}=0$ and therefore another approach is necessary.

\paragraph{Step~$1$: irreducibility of $f_s$ $(1\le s\le 2p+2q)$:}

Let us express $s = rpq + ip + jq$ in the sense of Lemma \ref{lem:rpqipjq}.
First we study the case where $r = 0$, $0 \le i \le q - 1$ and $j = 0, 1$.
If $j = 0$, we have $g_{2p} \neq 0$ and
\[
	g_{ip} = -g^k_{(i-1)p} + g^k_{(i-1)p} + g^{k^2}_{(i-2)p} = g^{k^2}_{(i-2)p}.
\]
Thus
\[
	g_{ip} = \begin{cases}
		1 & (i \text{ :odd}) \\
		2^{k^{i-2}} & (i \text{ :even})
	\end{cases}.
\]
Therefore $g_{ip} \ne 0$.
Let us study the case of $j = 1$.
For $i = 0, 1, 2$ we have
\[
	g_q = 1, \quad
	g_{q+p} = -1, \quad
	g_{q+2p} = 1.
\]
Here we have used Lemma \ref{lem:cpq} to obtain the value of $g_{q+2p}$.
For $i \ge 3$,
\begin{align*}
	g_{q+ip} &= \frac{-g^k_{q+(i-1)p} g^k_{ip} + \left( g^k_{q+(i-1)p} g^k_{(i-2)p} + g^k_{ip} g^{k^2}_{q+(i-2)p} \right) g^{k^2-1}_{(i-1)p}}{g^k_{(i-2)p}} \\
	&= \begin{cases}
		-g^k_{q+(i-1)p} + N_{i-2} \left( g^k_{q+(i-1)p} + g^{k^2}_{q+(i-2)p} \right) & (i \text{ :odd}) \\
		-(N_i/N_{i-2}) g^k_{q+(i-1)p} + \left( g^k_{q+(i-1)p} + (N_i/N_{i-2}) g^{k^2}_{q+(i-2)p} \right) & (i \text{ :even})
	\end{cases},
\end{align*}
where $N_i = 2^{k^{i-2}}$.
Since $g_{q+ip} \equiv 1 \mod 2$ we have $g_{q+ip} \ne 0$.

\noindent
Next let us prove the irreducibility of $f_{2q}$.
Since $g_{2q} = 0$, we cannot use the same argument as before using $g_{2q}$.
Let $h_s$ be the values of $f_s$ when we substitute
\[
	f_m = \begin{cases}
		2 & (m = -2p - 2q) \\
		-1 & (m = -2p) \\
		1 & (-2p - 2q \le m \le -1, m \ne -2p)
	\end{cases}
\]
in the initial variables of $f_s$ and take $a = b = 1$.
Then we have
\[
	h_0 = 0, \quad
	h_q = 1, \quad
	h_{2q} = 2 \ne 0.
\]
Thus $f_{2q}$ is irreducible.

\noindent
The proof for $g_{2q+p}, g_{2q+2p}, g_{3q} \ne 0$ is straightforward.
From Lemma \ref{lem:cpq} we have $g_{2q+p} = 1 \ne 0$.
Using $g_{2q} = 0$, we have $g_{2q+2p} = 2^{k^2} \ne 0$.
If $p \ge 4$, $g_{2q} = 0$ leads us to $g_{3q} = 1 \ne 0$.
If $p = 3$, since $3q = pq$, we can use the discussion in the next step.

\noindent
Next let us prove the case of $r=1$: i.e., $s = rpq + ip + jq = pq + ip + jq$.
Note that from the condition $s\le 2p+2q$, only the case $s=pq$ with $(p,q)=(3,4), (3,5)$ is possible.
Therefore $g_{(p-1)q} = g_{2q} = 0$ and thus $g_{pq} = g^k_{(q-1)p} g^{k^2}_q \ne 0$.

We have proved that $f_s$ is irreducible for $0 \le s \le 2p + 2q$.

\paragraph{Step~$2$: coprimeness of $f_s$ $(0 \le s  \le 2p + 2q)$:}
First let us present several properties concerning the variable $c_s^{(0)}$ in Lemma
\ref{lem:irredcs}.
\begin{Lemma} \label{lemstep56}
We have $c^{(0)}_{jp+i} < c^{(0)}_{jp}$ for any $j \in \mathbb{Z}_{\ge 0}$ and $0 < |i| \le p - 1$.
Moreover it holds that $c^{(0)}_{mp} \ge c^{(0)}_{(m-1)p}$ for any $m > 0$.
\end{Lemma}
\Proof

For the former inequality,
it is sufficient to show that, for a fixed $i$, $y_{m,n} := c^{(0)}_{mp+nq} - c^{(0)}_{mp+nq+i}$
satisfies $y_{m,0} > 0$.
Note that $y_{m,n}$ satisfies $y_{m+q,n} = y_{m,n+p}$ with the initial values
\[
	y_{-2,-2} = 1, \quad
	y_{m,-2} = y_{m,-1} = y_{-2,n} = y_{-1,n} = 0 \quad
	(-1 \le m \le q - 1, -1 \le n \le p - 1),
\]
and the recurrence relation for $y_{m,n}$ is
\[
	y_{m,n} = k \left( y_{m-1,n} + y_{m,n-1} - y_{m-2,n-1} - y_{m-1,n-2} \right) + y_{m-2,n-2}.
\]
If we define $d_{m,n} = y_{m,n} - y_{m-1,n-1}$ we have
$d_{m,n} = k ( d_{m-1,n} + d_{m,n-1} ) - d_{m-1,n-1}, \ d_{m+q,n} = d_{m,n+p}$,
with the initial values $d_{-1,-1} = -1, \
	d_{m,-1} = d_{-1,n} = 0 \
	(0 \le m \le q - 1, 0 \le n \le p - 1)$.
Therefore it is readily obtained that $y_{m,n}>0$ for $m,n \ge 0$.
If $m, n$ satisfy $-p \le m - n \le q$, we have
$y_{m,n} = \sum_{\ell} d_{m-\ell,n-\ell}$,
where the summation runs over $0 \le \ell \le \min(m, n)$.
Thus $y_{m,n} > 0$.
For a fixed $m \ge 0$,
let us take $m_0 \in \mathbb{Z}_{\ge 0}$ such that $-q \le m_0 (p + q) - m \le p$.
Then $y_{m,0} = y_{m - m_0 q, m_0 p}$ and $-p \le m - m_0 q - m_0 p \le q$
indicate that $y_{m,0} > 0$.
The latter inequality in Lemma \ref{lemstep56} is proved in a similar manner to the former one
by defining $e_{m,n} := c^{(0)}_{mp+nq} - c^{(0)}_{(m-1)p+(n-1)q}$, which satisfies the same recurrence as that for $d_{m,n}$, and by proving $d_{m,0} \ge k d_{m-1,0}$.
\qed

Now, let us prove that $f_s$ and $f_t$ are pairwise coprime for $0 \le s < t \le 2p + 2q$.
From (iii) of Lemma \ref{lem:irredcs}, it is sufficient to prove that $c_s \ne c_t$.
When we define
\[
	m_s = \max \left( c^{(0)}_s, \ldots, c^{(0)}_{s-p+1} \right),
\]
it is shown from Lemma \ref{lemstep56} that there exists $i \le j$ such that $m_s = c^{(0)}_{ip}$, $m_t = c^{(0)}_{jp}$.
If $i < j$ then we have $m_s < m_t$ and thus $c_s \ne c_t$.
If $i = j$ then again from Lemma \ref{lemstep56} we have $c_s \ne c_t$ since they attain their maxima in the distinct elements.

\paragraph{Step~$3$: irreducibility and coprimeness of $f_s$ $(2p+2q+1\le s)$:}
For $s \ge 2p + 2q + 1$, let us assume that $f_s$ is not irreducible.
Then $f_s$ must be factorized as $f_s = u f_r f_0$ using an invertible element $u$ and a subscript $r$ with $1 \le r \le 2p + 2q$.
From (ii) of Lemma \ref{lem:irredcs}, we have $c_s = c_r + c_0$, which contradicts the calculations in the previous steps.
It is readily obtained that $f_s$ and $f_t$ are pairwise coprime for every $0 \le s < t$
in the same manner as in the previous steps.
\qed
\subsection{Comments on Proposition \ref{lower_bd}}
Let us show two Lemmas on $\Lambda_{p,q}$.
\begin{Lemma} \label{Note_root}
We have $1<\Lambda_{p,q}<k$. Moreover,
$\Lambda_{p,q}$ is the largest absolute value among all the roots (including the imaginary ones) of \eqref{pq_roots}.
\end{Lemma}
\Proof
Let $f(\lambda):=\lambda^{p+q}-k(\lambda^p+\lambda^q)+1$, then $f(1)<0$ and $f(k)>0$ and thus $1<\Lambda_{p,q}$.
For $x \ge \Lambda_{p,q}$ we have
\begin{align*}
f'(x)&=x^{-1}\left\{ p(x^q-k)x^p+q(x^p-k)x^q\right\}\\
&\ge x^{-1}\left\{ p(\Lambda_{p,q}^q-k)\Lambda_{p,q}^p+q(\Lambda_{p,q}^p-k)\Lambda_{p,q}^q\right\}\\
&=x^{-1}\left\{  q(k\Lambda_{p,q}^p-1)+p(k\Lambda_{p,q}^q-1)  \right\}>0.
\end{align*}
Thus for $x>\Lambda_{p,q}$ we have $f(x)>0$ and thus $\Lambda_{p,q}<k$.
Assume that there exists a root $\lambda$ such that
$|\lambda|>\Lambda_{p,q}$, then $f(x)>0$ for $x:=|\lambda|$.
On the other hand, if we take $\lambda=x e^{-\sqrt{-1}\gamma}$, $f(\lambda)=0$ is equivalent to
\[
x^{p+q}-k(x^p e^{\sqrt{-1}q\gamma}+x^q e^{\sqrt{-1}p\gamma})+ e^{\sqrt{-1}(p+q)\gamma}=0.
\]
Let us prove that no $x,\,\gamma$ satisfy the above equality.
It is sufficient to show that, for fixed $1<u,\,y,\,z$ and $u+y<1+z$, we cannot find
$\theta,\,\phi$ that satisfy \eqref{thetaeq}:  
\begin{equation} \label{thetaeq}
z-(u e^{\sqrt{-1}\theta}+y e^{\sqrt{-1}\phi})+ e^{\sqrt{-1}(\theta+\phi)}=0.
\end{equation}
By taking the real part of \eqref{thetaeq} we have
\[
u+y-1+\cos(\theta +\phi)<z+\cos (\theta+\phi)=u\cos \theta +y \cos \phi.
\]
Thus
\[
(1-\cos \theta)+(1-\cos \phi) \le u(1-\cos \theta)+y(1-\cos \phi) <1-\cos (\theta+\phi),
\]
from which $0 \le (1-\cos \theta)(1-\cos \phi)<\sin \theta \sin \phi$
immediately follows.
On the other hand, by taking the imaginary part of \eqref{thetaeq} we have
\[
u\sin \theta +y \sin \phi =\sin (\theta+\phi),
\]
and thus
\[
(u-\cos \phi)\sin \theta +(y-\cos \theta)\sin \phi =0,
\]
which is impossible since $\sin \theta \sin \phi>0$.
\qed

\begin{Lemma}\label{Note_sol}
There exists a constant $c>0$ such that $d_m^* \ge c\Lambda_{p,q}^m$.
\end{Lemma}
\Proof
Since the degree $d_m^*$ satisfies $(d_{-2q-2p}^*,\cdots,d_{-1}^*)=(1,0,0,...,0)$ and
\[
d_{m}^*-k(d_{m-p}^*+d_{m-q}^*)+k(d_{m-2p-q}^*+d_{m-p-2q}^*)-d_{m-2p-2q}^*=0,
\]
there exist suitable constants $c_i \in \C$ such that $d_m^*=\sum_{i=1}^{2p+2q}c_i \lambda_i^{m+2p+2q}$, where $\{\lambda_i\}$ consists of $p+q$ roots of
\eqref{pq_roots} in addition to the $p+q$-th root of unity.
(Note that we omitted the case of multiple roots, however the discussion proceeds similarly to the simple roots.)
Let $\lambda_{2p+2q}=\Lambda_{p,q}$ and we prove that $c_{2p+2q} \ne 0$.
Let $\c:={}^t(c_1,c_2,...,c_{2p+2q})$，$\boldsymbol{e}_1:={}^t(1,0,0,...,0)$ and $\A$ be the square Vandermonde matrix generated by $\lambda_1,\lambda_2,...,\lambda_{2p+2q}$.
Then we have $\A\c=\boldsymbol{e}_1$.
From Cramer's rule we have
\[
c_{2p+2q}=-\frac{|\A_{1.2p+2q}|}{|\A|},
\]
where $|\A_{1.2p+2q}|$ is the $(1,2p+2q)$-first minor of $\A$.
The determinant of the Vandermonde matrix is nonzero, and $|\A_{1.2p+2q}|\neq 0$ is also satisfied, since $|\A_{1.2p+2q}|=\prod_{i=1}^{2p+2q-1}\lambda_i\times |B|$, where
$B$ is the square Vandermonde matrix generated by $\lambda_1,\lambda_2,...,\lambda_{2p+2q-1}$.
Thus $c_{2p+2q} \ne 0$.
Therefore we can choose $0<c \ll |c_{2p+2q}|$ so that $d_m^* \ge 0$ and $d_m^* \ge c\Lambda_{p,q}^m$.
\qed

\subsection{Comment on Lemma \ref{sequence}} \label{note_root2}
For a generic initial values there exists a constant $c>0$ in addition to $C>0$  such that
$c \Lambda_{p,q}^m \le |a_m| \le C \Lambda_{p,q}^m$.
In fact, the iterate $a_m$ is expressed as 
\[
a_m=\sum_{i=1}^{2p+2q}c_i \lambda_i^m \qquad (c_i \in \C),
\]
where $\lambda_1,\lambda_2,...,\lambda_{2p+2q} = \Lambda_{p,q}$ are the roots of the characteristic polynomial.
Since $\Lambda_{p,q}$ has the largest absolute value among the roots,  we can find 
$c>0$ such that $c \Lambda_{p,q}^m \le |a_m|$ as long as $c_{-2p-2q} \ne 0$.

\end{document}